\documentclass{iopjournal}

\usepackage{graphicx}
\usepackage{amsmath,amssymb,amsfonts}
\usepackage{lmodern}
\newcommand{\bfB}{{\bf{B}}}
\newcommand{\bfr}{{\bf{r}}}

\usepackage{caption}
\usepackage{subcaption}

\usepackage{natbib}
\begin{document}

\articletype{Paper} 

\title{Understanding and Quantifying Banana Coil Magnetic Fields and Forces for Enhanced Optimisation}

\author{Annika Zettl$^{1,*}$\orcid{0009-0003-8973-8545}, Tobias Schuett$^2$\orcid{0000-0001-5197-4348} and Sophia Henneberg$^{2}$\orcid{0000-0002-1949-7032}}

\affil{$^1$Max Planck Institute for Plasma Physics, Greifswald 17489, Germany}

\affil{$^2$Plasma Science and Fusion Centre, Massachusetts Institute of Technology, Cambridge, MA 02139, USA}

\affil{$^*$Author to whom any correspondence should be addressed.}

\email{annika.zettl@ipp.mpg.de}

\keywords{coil optimisation, stellarator, tokamak-stellarator hybrid, banana coils, quasi-axisymmetry}

\begin{abstract}
The optimised tokamak–stellarator hybrid concept \citep{henneberg_compact_2024} has the potential to combine tokamak and stellarator advantages to achieve magnetically confined fusion. These compact quasi-axisymmetric designs can have a low aspect ratio and large plasma volume, good particle confinement, and relatively simple coils. Previous work showed that such magnetic configurations can in principle be reproduced by a single type of non-planar `banana coil' alongside the conventional tokamak coilset \citep{henneberg_variety_2025}.
In this work, we optimise banana coils while also considering engineering constraints beyond simple geometric measures. We quantify the characteristic geometries of force-optimised banana coils and the magnetic fields they generate, and analyse the mechanisms by which forces may be reduced through optimisation.
\end{abstract}

\section{Introduction}
    Tokamaks and stellarator are two of the most widely studied concepts in magnetic confinement fusion. Their differences start with their shape, which in turn influences their plasma properties.
    While tokamaks are axisymmetric tori in shape, stellarators take on more complex geometries. Some advantageous features of tokamaks are that they are typically easier to construct, and have historically achieved better plasma confinement. Their simple geometry comes at the cost of requiring a large toroidal electric plasma current, which can lead to large scale plasma instabilities including disruptions.
    Stellarators need not rely upon a plasma current, and instead use intricately shaped magnetic fields to confine the plasma. On the other hand, shaped magnetic fields which are lacking a symmetry generally have poor confinement. Fortunately, the larger design space of stellarators offer a design freedom to optimise for desirable properties.  Notably, stellarators can be optimised to exhibit so-called `hidden symmetries' in the magnetic field strength that can ensure collisionless particle confinement \citep{helander_theory_2014, landreman_magnetic_2022, goodman2023constructing}.
    There have been different experiments to introduce a three-dimensional structure into an otherwise axisymmetric plasma, to combine the two concepts in certain ways, for example spherical stellarators \citep{moroz_spherical_1996, moroz_extreme_1998}, the Compact Toroidal Hybrid \citep{pandya_low_2015}, W7-A \citep{team_stabilization_1980,hirsch_major_2008}, or the tokastar machine \citep{yamazaki_tokastar_1985}, but none of these, have been transport-optimized to exhibit the hidden symmetries of stellarators.
    One of these is quasi-axisymmetry (QA), a concept that is closely related to the axisymmetry of tokamaks: in a special coordinate systems, the magnetic field strength is independent of the toroidal-like coordinate.  Stellarator equilibria exhibiting these symmetries share many properties with tokamaks, such as a net toroidal current or toroidally-closing contours of magnetic field strength. 
    It is therefore not surprising that the two concepts are continuously connected.
    In particular, one can numerically perturb axisymmetric boundaries while preserving quasi-axisymmetry, such that the resulting three-dimensional shaping adds rotational transform, as demonstrated numerically by \cite{boozer_stellarators_2008,
    ku_nonaxisymmetric_2009}, and
    more recently analytically for vacuum \cite{plunk_quasi-axisymmetric_2018} and at finite pressure and current \citep{plunk_perturbing_2020}. These tokamak configurations that are locally perturbed on the inboard side can also be obtained via optimisation in vacuum and at finite beta as presented by \cite{schuett_exploring_2024,schuett_optimization_2025}, targeting an increased external rotational transform, quasi-axisymmetry, and additional physics properties. Similarly shaped configurations in vacuum have also been found by \cite{nies2024exploration} and \cite{giuliani2025comprehensive}.
    Unlike previous quasi-axisymmetric designs \citep{nuhrenberg1994quasi, nelson2003NCSX, zarnstorff2001physics, henneberg2019properties, landreman_magnetic_2022}, these configurations have significant overlaps in their toroidal cross-sections, and can be similarly compact (low aspect ratio) as tokamaks \citep{henneberg_variety_2025, schuett_optimization_2025}.
    Moreover, \cite{henneberg_compact_2024, henneberg_variety_2025} have demonstrated that the required magnetic field for the perturbed QA equilibrium can be generated using the simple standard tokamak coils with only one additional type of coil, a so-called \textit{banana coil}.
    Owing to the large axisymmetric plasma volume contained within the stellarator boundary, combined with this coil concept, it is conceivable to design a single machine capable of operating either as a tokamak, or when switching on the additional banana coils, as a perturbed QA stellarator. \cite{henneberg_compact_2024} proposed this idea as a tokamak-stellarator hybrid, which we hereafter simply refer to as a \textit{Hybrid}.

\begin{figure}
    \centering
        \begin{subfigure}[b]{0.3\linewidth}
            \centering
            \includegraphics[width=\linewidth]{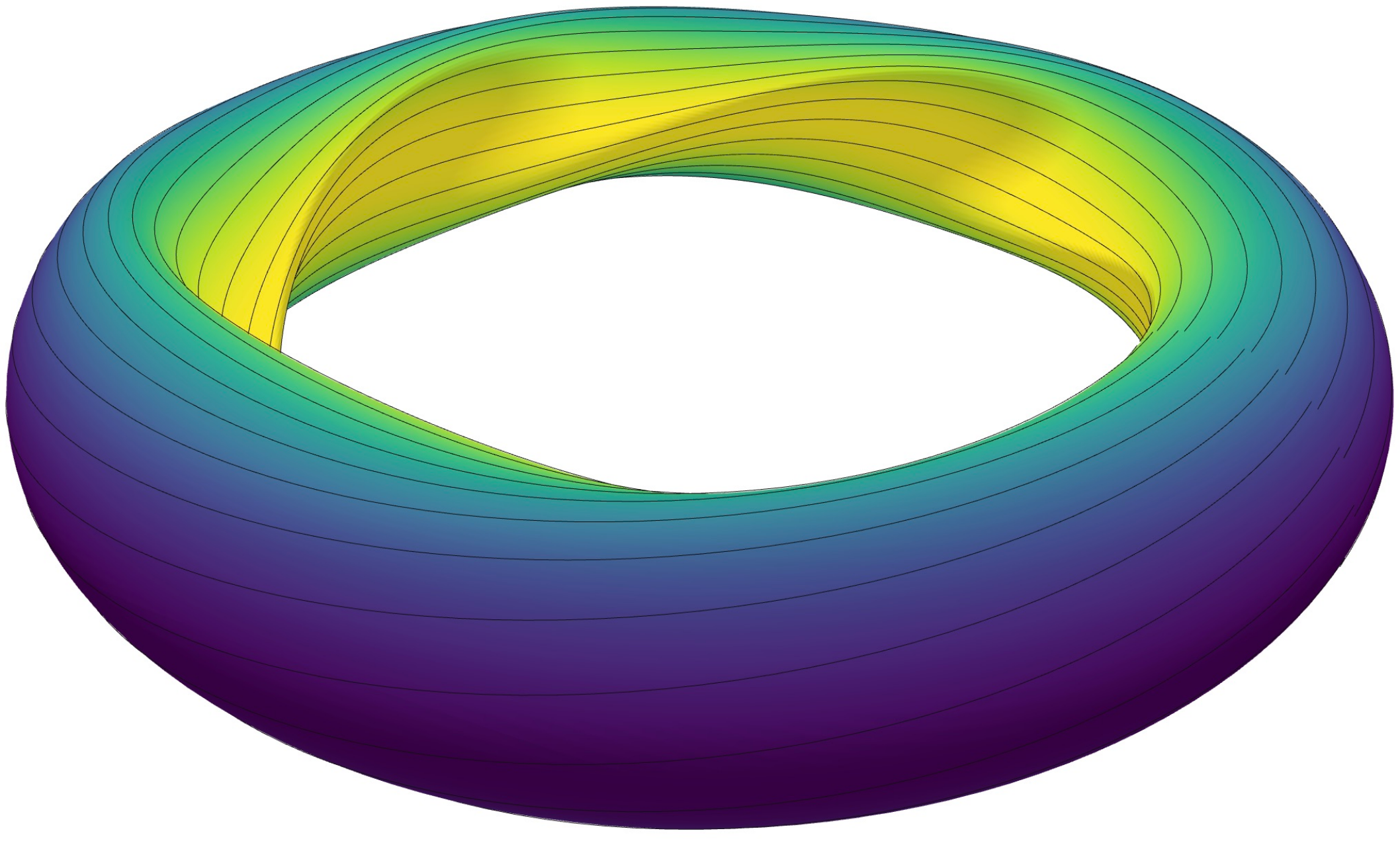}
            \caption{H1: $n_\text{fp}=4$, $\iota_\text{ext}=0.05$}
        \end{subfigure}
        \hspace{0.2cm}
        \begin{subfigure}[b]{0.3\linewidth}
            \centering
            \includegraphics[width=\linewidth]{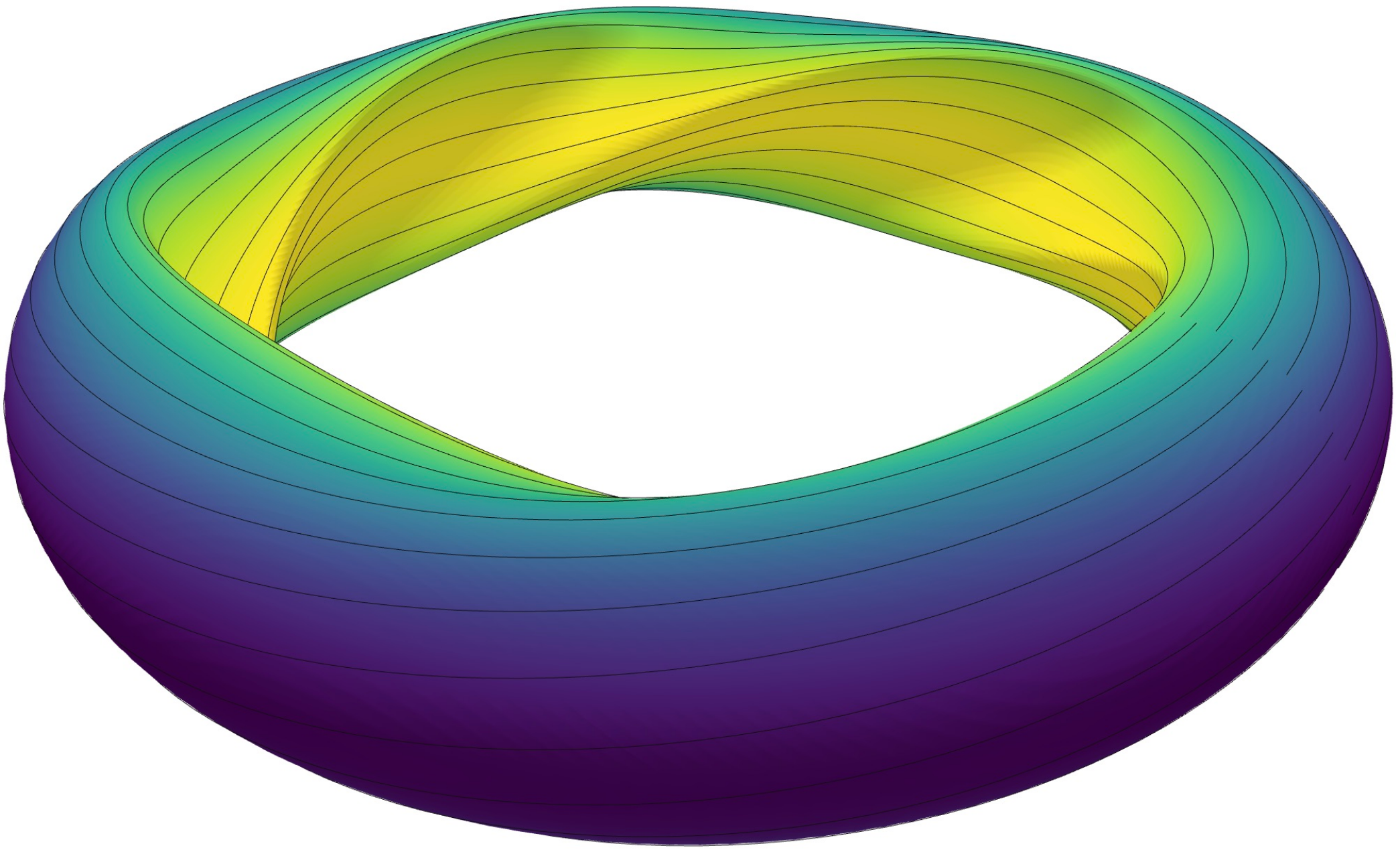}
            \caption{H2: $n_\text{fp}=4$, $\iota_\text{ext}=0.15$}
        \end{subfigure}
        \hspace{0.2cm}
        \begin{subfigure}[b]{0.3\linewidth}
            \centering
            \includegraphics[width=\linewidth]{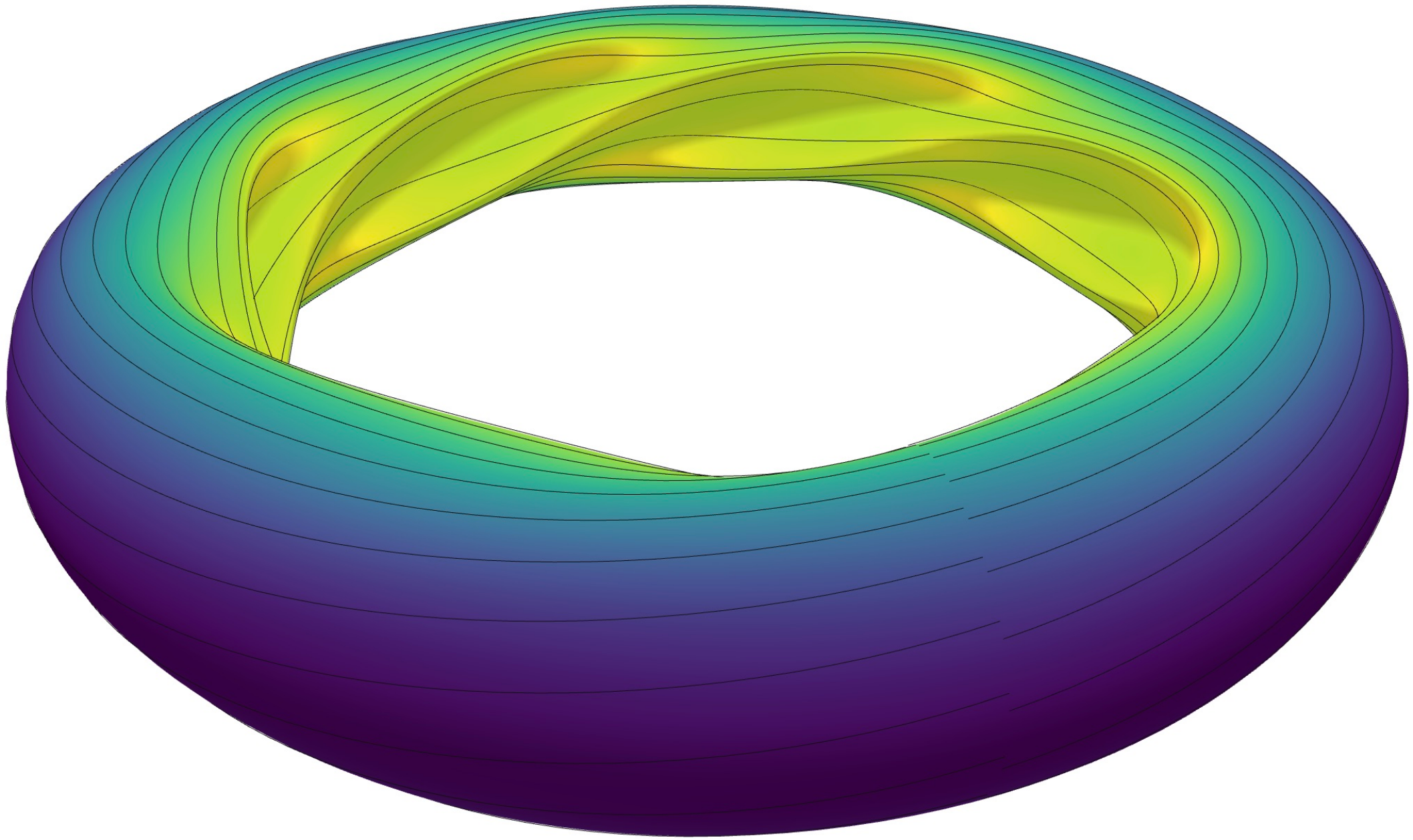}
            \caption{H3: $n_\text{fp}=6$, $\iota_\text{ext}=0.05$}
        \end{subfigure}
        \caption{The three different Hybrid equilibria for which we found coils in this work, labelled as H1, H2 and H3. The boundaries have the same aspect ratio $A=5$, but differ in the shape of the perturbations on their inboard sides, and in their number of field periods, $n_\text{fp}$, and external rotational transform, $\iota_\text{ext}$. The colour on the boundaries indicates the magnetic field strength, where blue indicates low fields ($B\gtrsim 3.1\textrm{T}$) and yellow high fields ($B\lesssim  4.3\textrm{T}$).}
        \label{fig:hybridequilibria}
    \end{figure}

    In a Hybrid equilibrium, the QA perturbations are located on the inboard side of the plasma, so that, when viewed from the outboard side, it still resembles the underlying tokamak. The perturbations themselves appear in each field period as helical \textit{grooves} framed by \textit{ridges}, and their exact shaping varies among Hybrids. These ridges, associated with a high principle curvature, tend to be straight, and also appear in magnetic fields of other QA configurations, as discussed in \cite{sengupta2026optical, sengupta2026compact}.
    In Figure \ref{fig:hybridequilibria} three of those equilibria are shown.
    Banana coils sit inside each of these grooves -- not linking with the plasma or the vessel -- and introduce the necessary 3D shaping in the magnetic field. They spiral up around a central $z$-axis and have an elongated, non-planar shape, which must be carefully optimised to accurately reproduce a plasma equilibrium.

    The complete Hybrid coil setup comprises the standard tokamak toroidal field (TF) and poloidal field (PF) coils, with the addition of a unique shape of a banana coil, which is periodically repeated on the inboard side of the plasma, respecting the number of field periods. The TF coils generate the dominant toroidal field, and the PF coils approximately cancel the vertical field that results from the plasma current. 

   In parallel with the development of the Hybrid concept, progress has also been made in numerical coil optimisation. In particular, a recently introduced method by Hurwitz and Landreman enables penalization of electromagnetic forces during coil optimisation \citep{hurwitz_electromagnetic_2025, hurwitz_efficient_2024, landreman2025efficient}.
    A concern regarding the feasibility of banana coils lies in the potentially tight turns, since they may lead to large electromagnetic forces.
    
    In this work, we apply this method to the design of banana coils, incorporating a range of geometric constraints to address engineering limitations. We then investigate how optimal banana coil configurations respond to these additional requirements.
    Central questions include the magnitude and spatial distribution of the resulting forces, as well as the extent to which they can be mitigated.
    
    The remainder of this paper is structured as follows. In Section \ref{sec:theory}, we first propose a simple mathematical description of the magnetic field generated by banana coils and subsequently analyse the electromagnetic forces arising from these models.
    To test these predictions, Section \ref{sec:stellsym} introduces the concept of self-stellarator symmetry to construct the desired banana coil geometries and outlines a method for generating a large set of force-optimised configurations.
    The results are presented in Section \ref{sec:results}, where we first examine the optimal coil geometries and then analyse the associated forces.
    Finally, we discuss the implications of our findings and draw conclusions in Section~\ref{sec:conclusion}.
        
\section{Mathematical Models for Electromagnetic Forces on Banana Coils} 
\label{sec:theory}
    \subsection{Candidate Models of Banana Coil Magnetic fields}
    \label{sec:analyticcoilmodels}
        For a simplified representation of a banana coil, we model it as comprising two elongated coil segments: one located closer to the plasma and the other farther away, connected at their ends by curved sections.
        At distances large compared to the conductor thickness, the coil can be approximated as a one-dimensional current filament, represented by a centreline curve $\mathbf{r}_c$ that carries the total current $I$. This approximation is adopted throughout this work to simplify both the model formulation and the subsequent coil optimisation.

        Two infinite straight wires that run parallel to each other and carry equal but opposite currents may be a fitting model for a banana coil.  
        At a distance $r$ from one infinitely long, straight current-carrying wire, the magnetic field is purely azimuthal,
        \begin{equation}
            \bfB_{\text{single}}(r) = \frac{\mu_0 I}{2\pi r} \hat{\mathbf{\phi}} .
            \label{eq:bfieldonestraight}
        \end{equation}
        Note that this relation also generally applies as an approximate description for the magnetic field of a coil section when the distance to that section is small compared to the radius of curvature.

        The magnetic field of two parallel straight wires with opposite currents in the $y$-direction and located at $x=\pm a$ on the $xy$-plane containing both wires, is 
        \begin{equation}
            \bfB_{\text{double},\parallel}(x) = \frac{\mu_0 I}{2\pi (x-a)} \hat{\mathbf{z}} + \frac{\mu_0 (-I)}{2\pi (x+a)} \hat{\mathbf{z}} = \frac{\mu_0 I}{\pi} \frac{a}{x^2-a^2}\hat{\mathbf{z}} \ ,
            \label{eq:bfieldtwostraight}
        \end{equation}
        and in the $yz$-plane at $x=0$, the centre of the two wires, we find
        \begin{equation}
            \bfB_{\text{double},\perp} (z) 
        =\frac{\mu_0 I}{2\pi \sqrt{a^2+z^2}} \cos(\theta_{-a}) \hat{\mathbf{z}} + \frac{\mu_0 (-I)}{2\pi \sqrt{a^2+z^2}} \cos(\theta_a)\hat{\mathbf{z}}
        = \frac{-\mu_0 I a}{\pi (a^2+z^2)} \hat{\mathbf{z}} ,
        \label{eq:bfieldtwostraightperp}
        \end{equation}
        where $\theta$ is the angle between the $x$-axis and the line joining the wire position $\pm a$ with the $z$ point, so that $\cos(\theta_{\pm a}) = \pm a/\sqrt{a^2+z^2}$.
        When sufficiently far away, the wire separation $d:=2a$ becomes negligible, and the field approaches a $1/r^2$ dependence on the distance $r$ to the coil.

        At larger distances, the detailed coil shape decreases even further in importance.
        The magnetic field of any coil can be approximated by that of a dipole field $\bfB_{\text{dipole}}$ with magnetic moment, ${\bf{m}}=I{\bf{A}}$, at distances much larger than the characteristic size $d$,
        \begin{equation}
            \bfB_{\text{dipole}}(r) = \frac{\mu_0}{4\pi} \left( \frac{3\hat{\mathbf{r}}({\bf{m}}\cdot\hat{\mathbf{r}}) - {\bf{m}}}{r^3}\right) ,
            \label{eq:idealdipole}
        \end{equation}
        where $\hat{\mathbf{r}}$ is the unit vector in the radial direction centred at the dipole.

        We analytically find three asymptotic regimes for candidate banana coil models: very close to a coil section ($r \ll d$)
        , where the coil effectively looks like a single straight wire, and its field scales as $B \sim I/r$. At intermediate distances ($r \sim d$), where the banana coil may be approximated by two long straight wires, the scaling becomes $B \sim I/r^2$. At distances large compared to the coil size ($r \gg d$), we expect the field to behave like that of a dipole, with $B \sim I/r^3$.
        Section \ref{sec:results} presents results to show whether these simple model can represent the magnetic field of banana coils and what candidate description can most accurately describe it. 

    \subsection{Contributions to Electromagnetic Forces on Banana Coils}
        A key consideration are the electromagnetic forces that act on the banana coils, as they need to be able to withstand the loads placed upon them. 
        Here, we discuss the origin of the forces acting on the banana coils. 
        To do so, we decompose the total magnetic field at the coil location into several contributions:
        \begin{enumerate}
            \item the field generated by the coil itself, $\bfB_\text{self}$, 
            \item the field due to the plasma current, $\bfB_\text{plasma}$, and 
            \item the fields from other coils divided into contributions arising from the toroidal field coils, $\bfB_\text{TF}$, poloidal field coils, $\bfB_\text{PF}$, and the mutual banana coils, $\bfB_\text{mutual}$.
        \end{enumerate}
        The force per unit length acting on a banana coil carrying a current $I_\text{banana}$ is given by Lorentz law,
        \begin{equation}
            \frac{\text{d}\mathbf{F}}{\text{d}l} = I_\text{banana} \bfr'_c \times (\bfB_\text{self} + \bfB_\text{plasma} + \bfB_\text{TF} + \bfB_\text{PF}+\bfB_\text{mutual}),
            \label{eq:fiveforcecontributions}
        \end{equation}
        where $\bfr_c'$ is the tangent vector to the banana coil curve.
        Based on this expression, one can reduce the magnitude of the pointwise electromagnetic forces on banana coils in four ways: by lowering the coil current, by minimizing the angle to the magnetic field of other coils, by avoiding the magnetic field altogether, or by cancellation, where the magnetic fields associated with the five terms in \eqref{eq:fiveforcecontributions} point in opposing directions such that the various force contributions subtract.

        \subsubsection{Self-Forces} 
        To approximate the self-forces acting on the two long sides of a banana coil, let us return to the two parallel straight wires with opposing currents.  
        Consider two wires separated by a distance $d$, aligned along the $x$-axis in Cartesian coordinates $(x,y,z)$. The configuration is such that the magnetic field produced by one wire at the location of the other then points in the $\hat{\mathbf{z}}$ direction, perpendicular to the plane containing both wires.
        The resulting force then pushes the two wires apart,
        \begin{equation}
            \frac{\text{d}\mathbf{F}_\text{self}}{\text{d}l} = I_\text{banana}\bfr_c' \times \bfB_\text{single} 
            = I_\text{banana}\hat{\mathbf{x}} \times \frac{\mu_0 (-I_\text{banana})}{2\pi d} \hat{\mathbf{z}} 
            = \frac{\mu_0 I^2_\text{banana}}{2\pi d} \hat{\mathbf{y}}.
        \end{equation} 
        If the two straight filaments are not parallel, but at an angle $\theta$ relative to each other, the force decreases as $\cos(\theta)$. A strategy to reduce this force, is therefore to increase the angle between the two filaments.
        Similarly, when a filament of a neighbouring banana coil is much closer than the coil width, the same relationship applies up to the sign of the current. The direction of the force inverts, and the two wires pull toward each other if the currents of the approaching coil segments point in the same direction.
        For banana coils that are far apart, the mutual magnetic force is $\sim I^2_\text{banana}/s^2$ for separations $s\gg a$, according to equations \eqref{eq:bfieldtwostraight} and \eqref{eq:bfieldtwostraightperp}.
        The self force dominates over the mutual force if the coil width is small compared to the coil-coil distance.
        
        A simple estimate of the self-forces at the tight endpoints is harder, however, the magnitude of the pointwise forces would be less than for parallel wires for similar separations and radius of curvature $d$. It should be noted here that because of the opposing direction of the forces at the turns, shear forces and stress may be significant.

        \subsubsection{Toroidal Field Forces}
        The toroidal field generated by the TF coils is inversely proportional to the radius $R$ in cylindrical coordinates, that is $\bfB_\text{TF} \approx (\mu_0 I_\text{encl}/2 \pi R) \ \hat{\boldsymbol{\phi}}$, where $I_\text{encl.}$ is the total current of all TF coils. 
        The force from this toroidal magnetic field on a banana coil is
        \begin{equation}
            \left|\frac{\text{d}\mathbf{F}_\text{TF}}{\text{d}l}\right| = I_\text{banana} \left|\bfr_c' \times \bfB_\text{TF}\right| =  \frac{\mu_0 I_\text{banana} I_\text{encl.}}{2\pi R}\sin(\theta) ,
            \label{eq:forceTF}
        \end{equation}
        where $\theta$ describes the angle of a banana coil filament to the toroidal field.
        To reduce forces, a banana coil filament should align horizontally with the toroidal field. This is most effective for the radially innermost coil sections, where the magnetic field is strongest.
        It is not feasible for the entire coil to align horizontally, since the upward inclination of the groove structure in Hybrid boundaries prescribes a necessary overall tilt angle for the coils. 
        In comparison to the endpoints, a typical banana coil is oriented at a moderate angle relative to the toroidal field over most of its length.
        Yet, at the endpoints the coil must inevitably bend through a right angle with respect to the toroidal field for its turn and is subjected to the full TF force, making this section susceptible to the largest forces anywhere on the coil. 
        Shifting the radius of the turning point $R_\text{turn}$ radially outward, into a region where the magnetic field is weaker, can reduce these forces.
        
        Placing the banana coils -- and their problematic turning points -- outside the TF coils avoids the toroidal field entirely. However, this advantage comes at a high cost: increasing the distance $r$ between the banana coil to the plasma requires a higher coil current. Using the double-wire approximation from section \ref{sec:analyticcoilmodels}, the scaling for the coil-plasma separation would be $I_\text{banana} \sim r^2$. Consequently, the self and mutual banana forces would scale as $\sim I_\text{banana}^2 \sim r^4$.
        Moving the coils radially inward additionally restricts the available room for coil-to-coil spacing and wide coils, which further amplifies both the mutual and self forces.
        When the banana coils are placed inside the TF coilset, the TF forces increase only as $ \sim I_\text{banana}/R_\text{turn} \sim r^2$ with the coil-plasma separation $r$ (assuming $R_\text{turn} = R_0-r \approx R_0$ for $R_0 \gg r$), though, the TF field here is very strong.
        Note that while a low current with low coil-plasma separations may seem like an attractive strategy, the proximity to the plasma gives rise to coil ripples in the boundary magnetic field that appear in an increased normal field error. Therefore, coils inside the TF coils should be optimised for minimal currents while still attaining a good quality in the reproduced equilibrium.
        
        \begin{figure}
            \centering
            \includegraphics[width=0.7\linewidth]{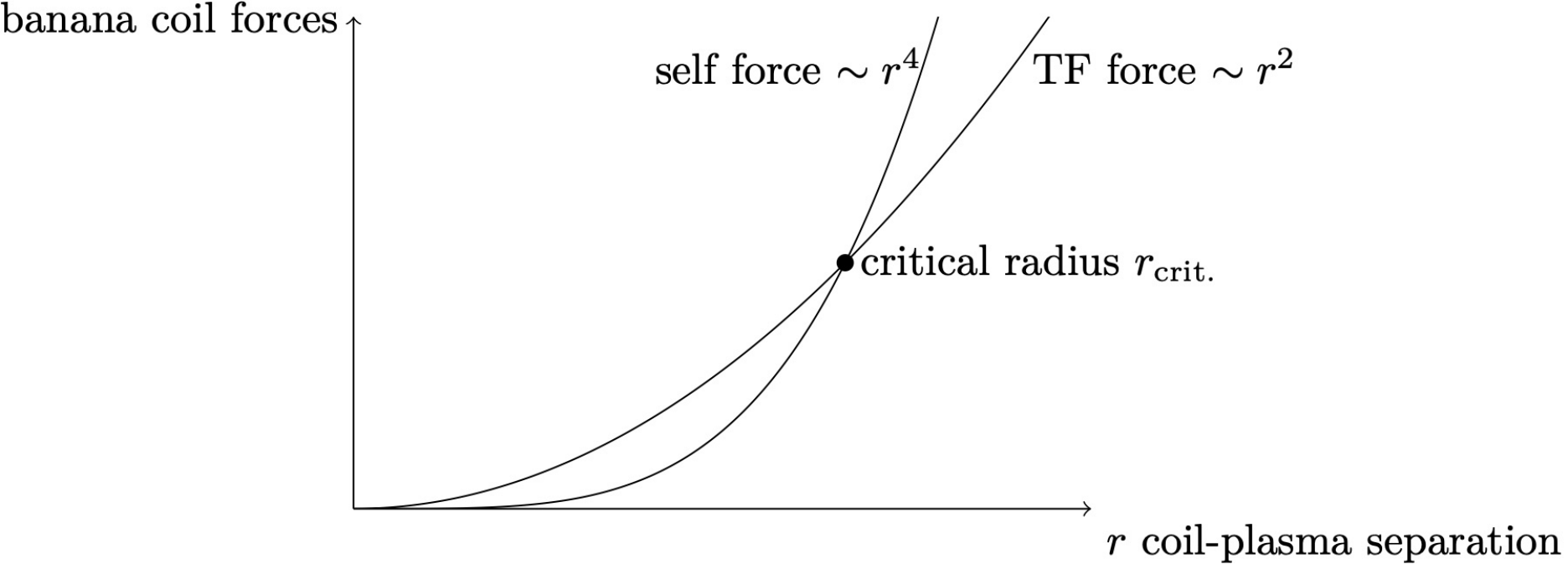}
            \caption{A schematic plot of the self and TF force on a banana coil as a function of its distance $r$ from the plasma. The critical radius, $r_\text{crit.}$, marks where the two force components equal.}
            \label{fig:criticalradius}
        \end{figure}
        
        The two relationships are schematically plotted in figure \ref{fig:criticalradius}. There is some \textit{critical radius}, denoted here as $r_\text{crit.}$, at which the self- and mutual forces equal those due to a toroidal field. This radius depends on the respective coil currents determining the field strengths and the detailed banana coil geometry. Beyond this critical radius, the self-forces exceed the toroidal field forces. If it lies outside the TF coil set, then placing the banana coils externally may be advantageous. Else, locating the banana coils inside of the TF coils results in lower overall forces.
        For an estimate of the critical radius, we find the banana coil current for which the self- and mutual forces from the straight section approximation become comparable to the toroidal forces from the turning points. For a typical width or coil-coil separation $d$, and a radial location of the coil turn $R_\text{turn}$, this occurs when
        \begin{equation}
            \frac{I_\text{encl.}}{R_\text{turn}} \approx \frac{I_\text{banana}}{d} .
            \label{eq:TFandselfaresimilar}
        \end{equation}
        Replacing $I_\text{banana}$ with the double-wire relation from equations \eqref{eq:bfieldtwostraight} and \eqref{eq:bfieldtwostraightperp} with $a\ll r$, that is $I_\text{banana} \sim 2 \pi B_\text{banana} r^2 /(\mu_0 d)$, where $B_\text{banana}$ is the targeted field magnitude generated by the banana coil at the plasma surface, we obtain
        \begin{equation}
            r_\text{crit.} \approx \sqrt{\frac{\mu_0 d^2 I_\text{encl.}}{2 \pi B_\text{banana} R_\text{turn} }} .
            \label{eq:rcrit}
        \end{equation}
        Note that this estimate decreases for narrower coils or larger radial locations of the turn.

        We now consider applying these considerations to the three equilibria used in this work (H1, H2, H3; see figure \ref{fig:hybridequilibria}). They have a major radius $R_\text{major} \approx 2.5\textrm{m}$ and on-axis total magnetic field strengths of around $B_0\approx 3.5T$ so that $I_\text{encl} \approx 45\textrm{MA}$, and banana coils could achieve $B_\text{banana} \approx 0.5\textrm{T}$ with a characteristic width of $d\approx 0.3\textrm{m}$. Assuming $R_\text{turn}\approx R_\text{major}$, the corresponding critical radius is $r_\text{crit.} \approx 0.8\textrm{m}$. This is inside typical TF coil sizes of a corresponding tokamak coilset ($\approx 1.5 \textrm{ m}$). 
        Thus, in this case, banana coils would be placed inside the TF coilset, and reducing the toroidal forces on the bend at the endpoint is best achieved by banana coil solutions with $R_\text{turn}$ as large as possible.

        \begin{figure}
            \centering
            \includegraphics[width=0.45\linewidth]{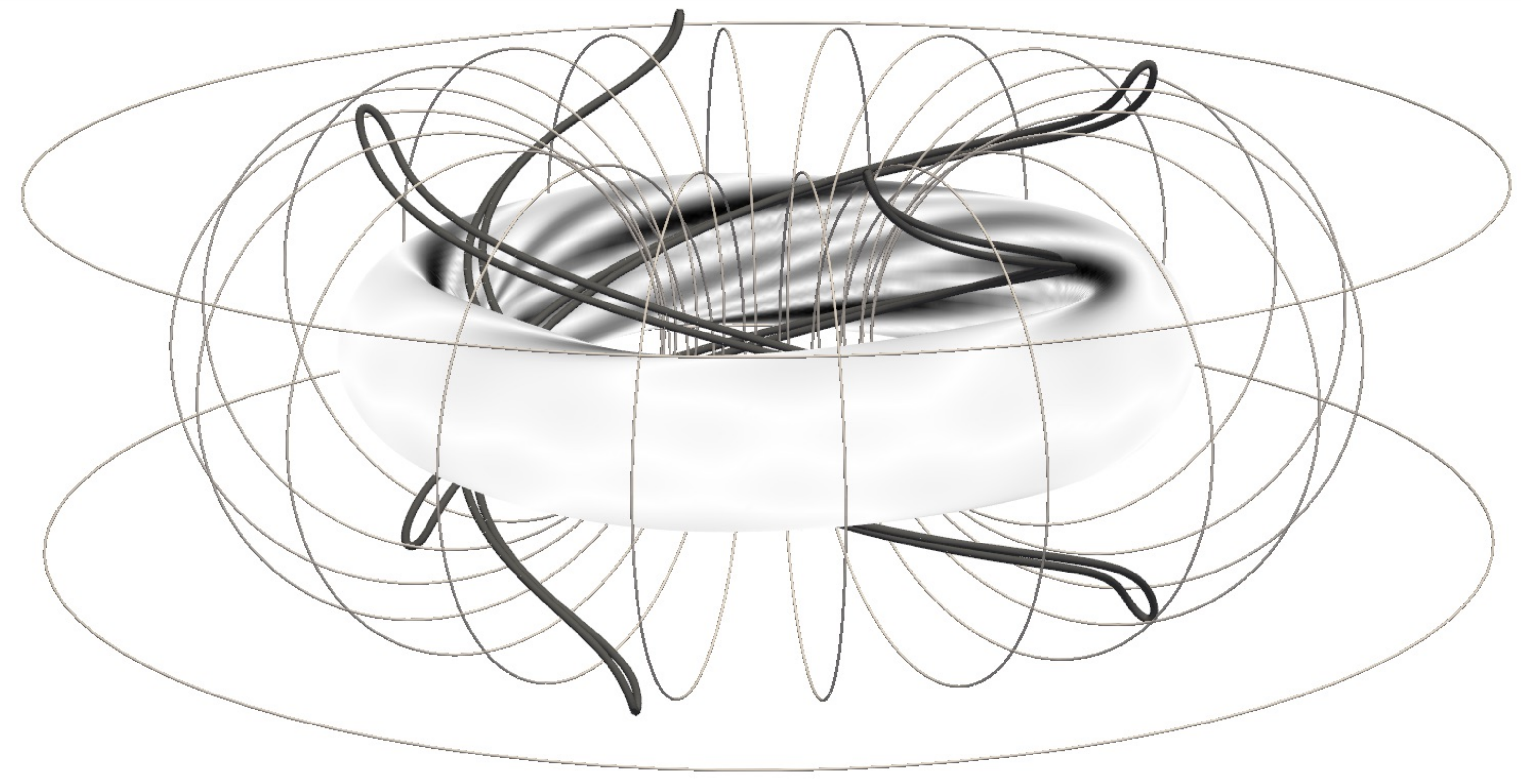}
            \caption{An example of ``escaping coils''. In this case, very long banana coils arrange themselves such that their endpoints pass between and extend beyond the toroidal field coils, such that their sharpest corners sit outside the toroidal field region.}
            \label{fig:escapingcoils}
        \end{figure}

        Alternatively, only the endpoints of banana coils could be placed outside the TF field. If the coil is positioned within the TF coil set but outside the plasma vessel, then, at least in principle, very long banana coils could be arranged such that their endpoints pass between and extend beyond the TF coils to make the turn outside the toroidal field, as shown in figure \ref{fig:escapingcoils}. (In this case, the coil ripple structure may actually be beneficial, as it can reduce the angle between the outgoing coil segment and the TF field.) 
        In any case, this strategy is only helpful if the forces arising from proximity to a TF coil does not dominate, and it might be challenging to construct in practice.

        \subsubsection{Poloidal and Vertical Field Forces} 
        The plasma current, when present, produces a poloidal magnetic field that decreases with distance from the plasma. Coils very close to the plasma would therefore experience the largest forces from this field. For typical tokamaks, the current in the plasma is about one to two orders of magnitude smaller than $I_\text{encl}$, and we would expect a similar scaling for the magnetic forces \citep{wesson, aymar2002iter, gruber_physics_1984}.
        The PF coils carry a current of similar size and opposite direction as the plasma current to generate a vertical field that opposes $\bfB_\text{plasma}$ on the inboard side of the plasma, and as a side effect, mitigates the forces due to the plasma current in this region.
        By their elongated nature, banana coils tend to be oriented at a large angle to both of these magnetic fields, unfortunately limiting the geometric strategies to avoid these forces.
        However, because of the much lower field strength, we expect the contribution from the plasma current and the PF coils to be less dominant, especially compared to the self- and TF forces.

        \subsubsection{Upper Bound on Forces}
        Neglecting the forces due to the poloidal and vertical field, the pointwise forces on typical banana coils arise mainly from sections of a coil that are at low radii or at a large angle to the toroidal field, as well as from the self-forces and mutual banana coils. The former arise especially near the endpoints, and the latter for narrow or closely spaced coils.
        We estimate the forces based on equations \eqref{eq:forceTF} and \eqref{eq:TFandselfaresimilar} to be bounded by
        \begin{equation}
            \left|\frac{\text{d}\mathbf{F}}{\text{d}l}\right| \lesssim \frac{\mu_0 I_\text{banana}}{ 2\pi } \left(\frac{I_\text{encl.}}{R_{\min}} + \frac{I_\text{banana}}{d_{\min}} \right),
            \label{eq:forcebounds}
        \end{equation}
        where $R_{\min}$ and $d_{\min}$ are lower estimates for the radial locations with perpendicular coil sections and the coil separations, width and coil-coil distance, respectively.

\section{Enforcing Stellarator Symmetry}
\label{sec:stellsym}

    In most existing stellarator designs, the plasma boundary is stellarator symmetric: configurations looks the same when looked at from the top and bottom \citep{dewar_stellarator_1998}.
    For scalar fields, in a cylindrical coordinate system $(R,\phi,z)$ stellarator symmetry is a rotational symmetry about an axis at $\phi=0$, such that
    \begin{equation}
        f(R,\phi,z) = f(R,-\phi, -z) .
        \label{eq:stellsym}
    \end{equation}
    For vector fields, $\mathbf{V}(R,\phi,z)=(V_R, V_\phi, V_z)|_{(R,\phi,z)}$, such as the magnetic field or a current distribution, this symmetry requires
    \begin{equation}
        (V_R, V_\phi, V_z)|_{(R,\phi,z)} = (-V_R, V_\phi, V_z)|_{(R,-\phi,-z)}.
        \label{eq:stellsym_vf}
    \end{equation}
    
    In particular, this implies that the electromagnetic coils need to be themselves stellarator symmetric.
    The standard approach for constructing modular coils that respect this symmetry
    begins with a set of $n$ distinct modular coils within a single half field period of the boundary. These coils are geometrically extended to the neighbouring half period by rotating the coils by an angle of $\pi$ about the stellarator symmetry axis. Under this transformation, the geometric shape of each coil is mirrored and flipped. The direction of the current is assigned according to the vector field definition of stellarator symmetry, 
    implying that it retains its poloidal sense of rotation. After this operation, there are $2n$ coils spanning one full field period. Repeating this full set across all $N_{fp}$ field periods amounts to a total of $2n\cdot N_{fp}$ coils. 
    
\begin{figure}
        \centering
        \begin{subfigure}[t]{0.25\linewidth}
        \includegraphics[width=\linewidth]{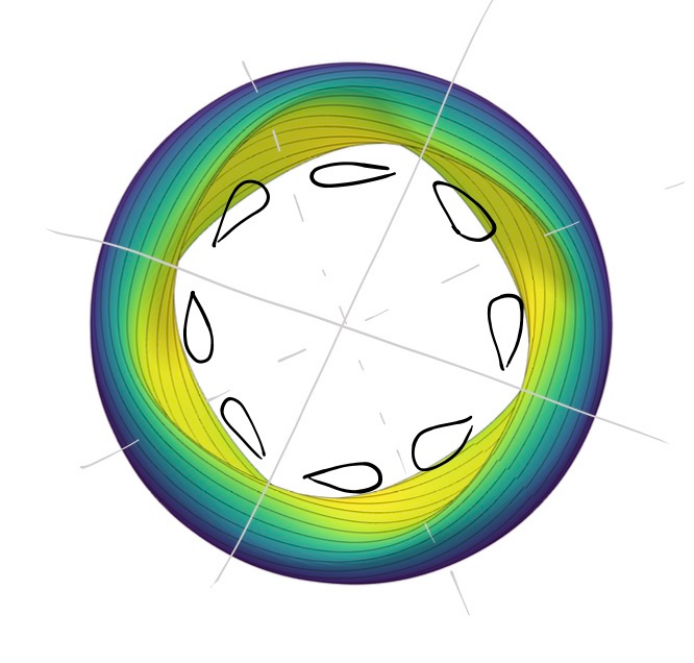}
        \caption{Mini-banana coils}
        \label{fig:minibananassketch}
        \end{subfigure}
        \begin{subfigure}[t]{0.25\linewidth}
        \includegraphics[width=\linewidth]{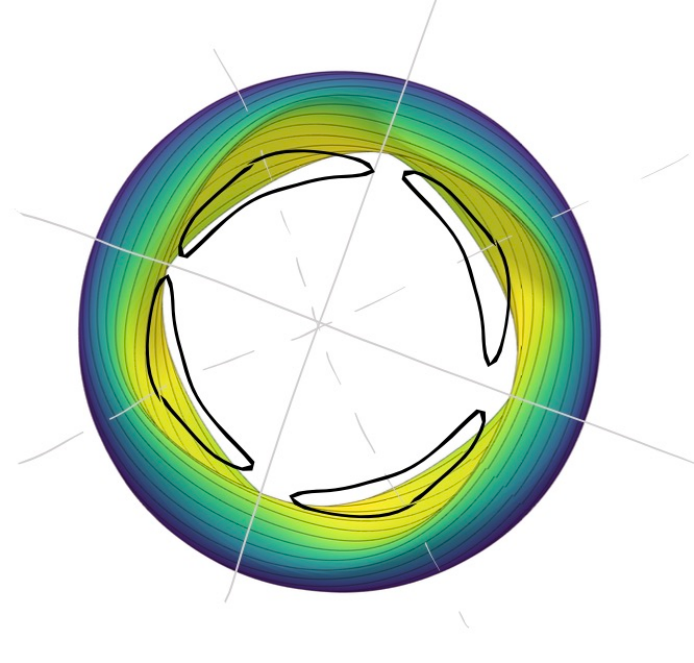}
        \caption{Full banana coils}
        \label{fig:selfstellsymsketch}
        \end{subfigure}
        \caption{A sketch of the two options to place non-linked coils respecting stellarator symmetry. In (a) are so-called mini-banana coils, where one coil per field period is placed that is mirrored and repeated for every field period, and (b) illustrates self-stellarator symmetric banana coils, with only one coil per field period.}
        \label{fig:stellsymillustration}
    \end{figure}

    If we were to apply the standard approach to banana coils in Hybrids, we would end up with at least two coils per field period, that is two coils per groove as sketched out in figure \ref{fig:minibananassketch}.
    To place one coil per field period, as illustrated in figure \ref{fig:selfstellsymsketch}, the banana coil must itself fulfil stellarator symmetry, and to achieve this, we introduce the concept of \textit{self-stellarator symmetry} for coils.
    This imposes a geometric restriction on the curve tracing out the coil centreline.
    For convenient notation, we centre a cylindrical coordinate system $(R,\phi,Z)$ along a stellarator symmetry axis at $\phi=0, Z=0$. 
    The coil centreline is a closed, connected curve with a parametrisation
    \begin{equation}
        \mathbf{x}(\theta) = (R(\theta), \phi(\theta), Z(\theta))
    \end{equation}
    for an angle $\theta \in [0, 2\pi)$ that uniquely defines the point on the curve. Self-stellarator symmetry requires that for any point on the curve $\mathbf{x}_1  = \mathbf{x}(\theta_1)$, there exists a point $\mathbf{x}_2  = \mathbf{x}(\theta_2)$ that satisfies
    \begin{equation}
        \mathbf{x}_2 = \mathbb{R}_1(\pi) \cdot \mathbf{x}_1,
    \end{equation}
    where $\mathbb{R}_1(\vartheta)$ is the rotation matrix about the $R$-axis by an angle $\vartheta$. 
    In terms of the parametrisation, self-stellarator symmetries means
    \begin{equation}
        (R(\theta_2), \phi(\theta_2), Z(\theta_2)) = (R(\theta_1), - \phi(\theta_1), - Z(\theta_1)) .
        \label{eq:selfstellsym_cyl}
    \end{equation}
            
    Points on the symmetry axis remain fixed under the rotation and therefore satisfy the condition trivially with themselves. Every other point forms a pair with its rotated counterpart. Consequently, the parameter space can be halved while still uniquely describing the closed curve. In other words, we can just choose a parametrisation to reflect this behaviour, so that $\mathbf{x}(0) = (R(0),0,0)$ is on the symmetry axis and $\theta_1 = -\theta_2$.
    In a Cartesian representation, ${\bf{x}}(\theta) =(x(\theta),y(\theta),z(\theta))$, applying the coordinate transformation $(R,\phi,Z) \mapsto (x,y,z) = (R\cos\phi, R \sin\phi, Z)$, yields an equivalent condition for self-stellarator symmetry,
    \begin{equation}
        (x_2, y_2, z_2) = \mathbf{x} (\theta_2) = \mathbb{R}_1(\pi) \cdot \mathbf{x}(\theta_1) = (x_1, -y_1, - z_1) .
    \end{equation}
    Note that the symmetry relation is even in sign in respect to $x$, and odd in $y$ and $z$. Any function parametrising these components must mirror this property. 
    For a Fourier representation of the coil curves in Cartesian coordinates, if $x(\theta)$ is to be even, it must be made up only of a constant and cosine terms. Conversely, the components $y(\theta)$ and $z(\theta)$ are a sum of sine terms. The parametrisation of the coil curves then reduces to

    \begin{equation}
    \left. \begin{array}{l}
    \displaystyle
        x(\theta) = x_0 + \sum x_{m}\cos(m\theta) \ , \\
        y(\theta)  = \sum y_{m} \sin(m\theta) \ , \\
        z(\theta)  = \sum z_{m} \sin(m\theta) \ .
        \label{eq:selfstellsym_Fourier}
    \end{array} \right\}
\end{equation}

    As a consequence, self–stellarator symmetric banana coils can never be entirely flat to an equilibrium boundary such that its open side faces the plasma.
    Instead, the coil curve must cross the stellarator symmetry axis at least twice, since neither $y$ nor $z$ in \eqref{eq:selfstellsym_Fourier} admits a constant term.
    At the coil's midpoint, it is therefore necessarily oriented laterally relative to the plasma boundary.
    The rest of the banana coil, however, is free to twist and rotate so that both endpoints could lie flat against the boundary. These twists and bends must of course be mirrored symmetrically on the opposite sides, which gives banana coils their distinctive look. 
    
    While self-stellarator symmetry has been thought of especially with banana coils in mind, its application is not limited to them. Modular coils can also be self-stellarator symmetric and allow for an odd number of coils per field period, or for a careful placement of a coil centred at the critical stellarator bean cross-section, which occurs at a symmetry axes.

\section{Methods to optimise and evaluate banana coils}
\label{sec:methods}
\subsection{Automatising the Optimisation} 

        We developed an automated scheme to optimise banana coils in the Hybrid coil configuration consisting of TF, PF and banana coils. This approach involves running many optimisations, where each time the exact target function, more specifically the corresponding weights and threshold values, varies randomly within some margins. The coils were represented as filaments, and the optimisations were implemented within the \texttt{simsopt} framework \citep{Landremansimsopt2021}. 
        We applied this technique to three similar optimised Hybrid equilibria (H1, H2, and H3, see table \ref{tab:configurationoverview} and figure \ref{fig:hybridequilibria}) that were computed using \texttt{VMEC} \citep{hirshman_steepest-descent_1983}.
        These vary in their shape, notably in their number of field periods and their external iota.

    \begin{table}
        \centering
        \begin{tabular}{l|c|c|c|}
             plasma configuration & H1 &H2 & H3 \\
             number of field periods $n_\text{fp}$ & $4$ & $4$ & $6$ \\
             external iota $\iota_{\text{ext}}$ & $0.05$ & $0.15$ & $0.05$ \\
             aspect ratio $A$                   & \multicolumn{3}{c|}{$5$} \\
             major radius $R_\text{major}$      & \multicolumn{3}{c|}{$2.5\mathrm{m}$} \\
             plasma current $I_\text{plasma}$   & \multicolumn{3}{c|}{$800\mathrm{kA}$ }\\
             on-axis magnetic field $B_0$       & \multicolumn{3}{c|}{$3.5\mathrm{T}$} \\
             volume averaged plasma $\beta$     & \multicolumn{3}{c|}{$1.2\%$}
        \end{tabular}
        \caption{Overview of several basic parameters of the three different Hybrid plasma equilibria, labelled as H1, H2 and H3, for which coils were optimised in this work.}
        \label{tab:configurationoverview}
    \end{table}
        
        The tokamak coils were not part of the optimisation, and were chosen a priori in the following way: the TF coils were centred at a major radius $R_{0,\text{TF}}=2.44\text{m}$ with coil radius $R_{1,\text{TF}}= 1.7\text{m}$. The total number of TF coils was $n_\text{TF}=20$ for $n_\text{fp}=4$ or $n_\text{TF}=18$ for $n_\text{fp}=6$.
        Two PF coils were placed at equal distances above and below the equilibrium with identical current, $I_\text{PF}$, optimised once for each of the three equilibria before adding any other coils.
        The banana coil optimisations started from a cold-start with the initial shape being a self stellarator-symmetric ellipse leaning into the direction of the Hybrid groove. 
        The banana coil current was randomly assigned a value between $0.5\%$ to $6\%$ of the total enclosed current ($225\text{kA}<I_\text{banana}<2.75\text{MA}$), and was held fixed throughout the optimisation. The degrees of freedom in the optimisations only consisted of the coefficients of the banana coil curve's Fourier representation, which comprised $5$ modes and were restricted by self-stellarator symmetry, amounting to a total of $16$ free parameters.
        
        The complete target function was
        \begin{equation*}
            f = f_\Phi + f_\text{L} + f_\kappa + f_\text{cc} + f_\text{F} + f_\text{wall} + f_\text{link},
        \end{equation*}
        with the different objectives defined as
        \begin{align}
        &f_\Phi = \int_{\mathcal{S}_\text{plasma}} \left(\bfB_\text{ext}\cdot \hat{\mathbf{n}} + \bfB_\text{plasma} \cdot \hat{\mathbf{n}}\right)^2 \text{d}s 
        \label{eq:targetflux} \ ,\\
        &f_\text{L} = \omega_\text{L} \max\{ L-L_*,\ 0\}^2 
        \label{eq:targetlength} \ ,\\
        &f_{\kappa} = \omega_{\kappa} \int_{\mathcal{C}}\max\{\kappa-\kappa_*,\ 0\}^2 \text{d}\ell 
        \label{eq:targetcurv}\  ,\\
        &f_\text{cc} = \omega_\text{cc}\sum_{i=1}^{n_\text{coils}} \int_{\mathcal{C}_0}\int_{\mathcal{C}_i}\max\{d_{\min}-|\mathbf{r_0}-\mathbf{r}_i|,\ 0\}^2 \text{d}\ell _0 \text{d}\ell_i 
        \label{eq:targetccdist} \ ,\\
        &f_\text{F}  = \omega_{\text{F}} \int_{\mathcal{C}} \max \{F  - F_*, 0\}^2 \text{d}\ell 
        \label{eq:targetforce} \ , \\
        &f_\text{wall} = \omega_\text{cs} \int_{\mathcal{C}}\int_{\mathcal{S}_\text{wall}}\max\{d_{\min}-|\mathbf{r}-\mathbf{s}|,\ 0\}^2 \text{d}\ell \text{d}s
        \label{eq:targetwall} \ ,\\
        &f_{\text{link}} = \sum_{i=1}^{n_\text{coils}} \frac{1}{4\pi} \left| \int_{\mathcal{C}_0}\int_{\mathcal{C}_i}\frac{\textbf{r}_0 - \textbf{r}_i}{|\textbf{r}_0 - \textbf{r}_i|^3} (d\textbf{r}_0 \times d\textbf{r}_i) \right|
        \label{eq:targetlink}  \ ,
        \end{align}
        where $\mathcal{C}_i$ is the curve along the i-th banana coil, $\bfr$ a location on this curve, and $\mathbf{s}$ a location on the wall surface. A $*$ subscript denotes threshold values that specify up to which value a quantity is penalised, and $\omega$ represents the associated weight that scales each objective.
        The target function therefore included, alongside the squared flux objective $f_\Phi$ penalising the normal field error $|\bfB\cdot \hat{\mathbf{n}}|$, $f_L$: the banana coil length $L$, $f_\kappa$: curvature $\kappa$ term, $f_\text{cc}$: coil-coil distance term, and $f_\text{F}$: the magnitude of the pointwise forces $F$ in vacuum on a banana coils using the regularisation with a circular conductor cross-section from \cite{hurwitz_efficient_2024}. 
        In addition, we introduce an objective $f_\text{wall}$ penalizing coils that approach a surface $\mathcal{S}_\text{wall}$ with the same major and minor radius as the TF coils to prevent banana coil solutions with escaping endpoints outside the TF coilset, like those in figure \ref{fig:escapingcoils}, and an objective to penalise interlinked coils ($f_\text{link}$). 
        The force objective employed a threshold value $F_*$ of zero. To examine the effect of this threshold, an additional set of coils was optimised for the H1 equilibrium alone, this time using a threshold of $1.75\textrm{T}\times I_\text{banana}$.

    \subsection{Coil Descriptors Beyond the Target Function} 
    \label{sec:beyondtarget}
    
        To describe the shape of a banana coil, we need to capture various geometric characteristics beyond those used in the target function, such as the length of the coil and its curvature. Understanding these characteristics is useful for classifying various features of banana coils, allowing us to learn how optimal coil solutions tend to behave. Since banana coils can take on many different shapes, it is not obvious how to define their key features in the best way. 
        In this section, we present three such measures which try to address this problem: the \textit{central coil width}, a \textit{net twist angle}, and the \textit{central curve-plasma distance}.
        \begin{figure}
            \centering
            \includegraphics[width=0.4\linewidth]{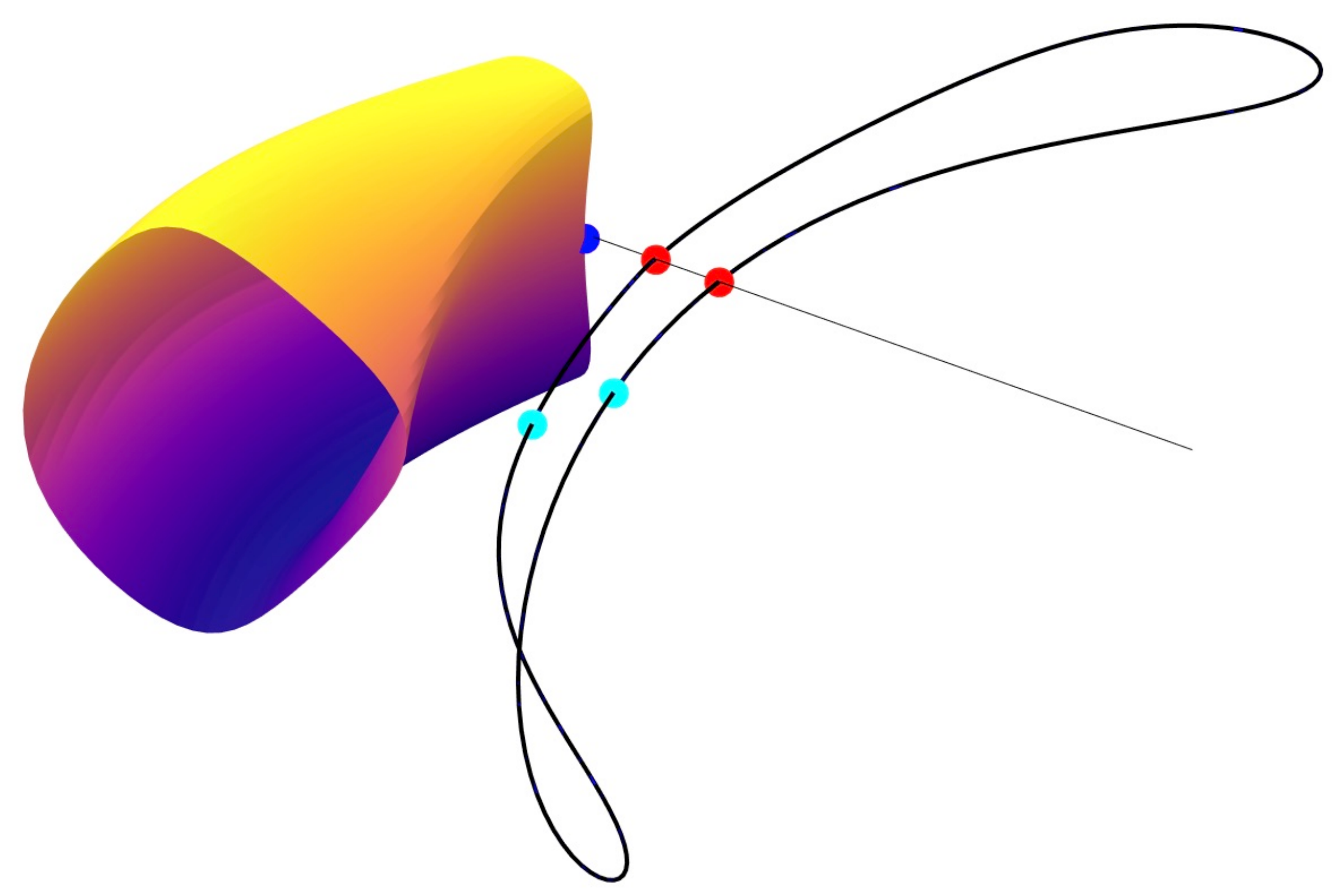}
            \caption{A diagram illustrating a banana coil beside a plasma boundary, with red dots showing the two central points, $R_\text{c}$ and $R_\text{d}$, a blue dot indicating the central surface point, $R_\text{s}$, and cyan points marking the coil locations at $\phi=\pi/2n_\text{fp}$, used for defining the net twist angle.}
            \label{fig:geodescriptors}
        \end{figure}

        For simplicity, we adopt a cylindrical coordinate system $(R,\phi,z)$, such that the stellarator symmetry axis of the banana coil of interest is in the $\hat{\mathbf{R}}$ direction at $\phi=z=0$.
        As shown in figure \ref{fig:geodescriptors}, the shape of the coil near the symmetry axis, where the two identical halves of a self-stellarator symmetric coil meet, and which we refer to as the \textit{centre}, can act as an indicator for the shape of the entire coil. For example, when the coil filaments are at a large angle to each other near the centre, this twist tends to extend throughout the entire coil shape. 
        The coil will intersect the stellarator symmetry axis at the centre in two places. Naturally, one will lie closer to the plasma boundary than the other, and we denote these by $R_\text{c}$ and $R_\text{d}$ respectively.
        The width $w$ of a banana coil refers to the separation between the two long sides. We define the \textit{coil width} at the centre as the distance between the points $R_\text{c}$ and $R_\text{d}$, that is $w=R_\text{d}-R_\text{c}$. 

        To get a measure for how parallel or twisted a coil is, we project the coil curve radially outward and find the angle between the two crossing lines at the coil centre. Typically, the filament closer to the plasma boundary aligns with the slope of the Hybrid groove, whereas the filament farther away may vary more in its direction.
        To determine this \textit{net twist angle} $\theta_{\text{twist}}$, we first compute the slopes of the two lines.
        To take a ``net slope'' over a finite segment of the coil we follow the curve, starting at the centre point at $\phi=0$, until we reach the point on the curve that sits at toroidal angle $\phi=\Delta\phi=\pi/2n_\text{fp}$. We then draw a straight line between this point and the centre point at $\phi=0$.
        The slope of this straight line in the $\phi z$-plane, $m_\text{c,d} =  \Delta z/\Delta \phi$,
        is the ``net slope'', and the angle between the two slopes $m_\text{c}$ and $m_\text{d}$ is
        \begin{equation}
            \theta_{\text{twist}} = \arctan\left(\frac{m_\text{c}-m_\text{d}}{1+m_\text{c} m_\text{d}}\right).
            \label{eq:twist}
        \end{equation}

        Last, we define the \textit{central coil–surface distance} as a measure of the separation between the coil and the plasma boundary. This is the distance between the point $R_\text{c}$ on the coil curve and a third, newly defined point $R_\text{s}$, located on the inboard side of the plasma boundary on the stellarator symmetry axis, and is called the \textit{central surface point}. 
        
        Note that these measures should only be regarded as proxies for the true geometry.
        The central coil width and coil-surface distance both ignore any variations in the spacing over the length of a coil, but still give insightful estimates for these quantities.
        The net twist angle is unable to capture certain features. For example, a wave-like structure over the averaging length, with large variations in slope, could yield the same average as a straight coil segment if the sampling points after $\Delta \phi$ and $\Delta z$ coincide spatially. 
        To detect wavy structures where the coil bends so sharply that its centre section forms an S shape and the curve goes in the reverse $\hat{\mathbf{\phi}}$ direction before reaching $\phi = \Delta \phi$ we additionally track the sign of the $\phi$-component of the tangent vector at the centre point.

    \subsection{Processing Automatically Optimised Coils}
    Among the automatically optimised coils can be unacceptable solutions that have high field errors or display undesirable shapes, such as extra loops, pronounced curls, self-intersections, or multiple twists near their endpoints.  
    We therefore applied filtering criteria to discard optimisation outcomes that are infeasible.
    First, to ensure the coils reproduce the target equilibrium to a prescribed accuracy, we imposed that the maximum field error, $|\mathbf{b}\cdot\hat{\mathbf{n}}|_\text{max}$, where is the normalized magnetic field $\mathbf{b} := \bfB/B$, must remain below $12\%$.  
    A further geometric criterion was needed to retain only those  optimised coils that have the desired elongated, simple shape.  
    As a preliminary, coarse filter to exclude coils with overly complex, looped geometries, we projected each coil curve onto a plane, perpendicular to the stellarator symmetry axis in a Cartesian coordinate frame, and required that the resulting projected curve have no more than three self-intersections. This criterion allows for the inevitable crossing at the symmetry axis and, in addition, one further intersection that can occur due to a twist near the coil ends to follow the curved plasma boundary.
    We additionally rejected coil configurations exhibiting an S-shaped turn at the more distant point to the boundary, $R_\text{d}$, and required that the $\phi z$-projected slope of the banana coils at the closer point $R_\text{c}$ have the same sign as the Hybrid groove slope. This ensures that the coils approximately follow the groove direction and suppresses kinks that may emerge from artefacts of higher Fourier modes. Other features, that render coils unfeasible -- already when accounting for their finite thickness -- and break the assumptions underlying the regularisation of self-forces, are extreme curvatures ($\kappa>80\textrm{m}^{-1}$), small coil-coil distances ($d_\text{cc}<5\mathrm{cm}$), and close coil-plasma distances ($d_\text{cs}<5\mathrm{cm}$). Coil solutions exhibiting any of these features were therefore also discarded. The threshold values were chosen deliberately generous so as not to exclude too many solutions. 
    Individual coils whose properties strongly deviated from those of the remaining set were examined manually and removed when necessary.

\section{Results on force-optimised banana coils} \label{sec:results}
    Applying these methods, a large set of banana coil configurations was automatically optimised for the three Hybrid equilibria, H1, H2 and H3, of which a total of $5093$ satisfied the filtering criteria (see table \ref{tab:fluxnumbers}). This section primarily presents the results for the H1 equilibrium, alongside selected figures from the other two optimisation sets. Additional plots are provided in appendix \ref{app:extradata}. 
    Note that an overlap between the optimised coil solutions for the first equilibrium—with and without a force threshold—is unavoidable. For sufficiently small force weights $\omega_\text{F}$, the contribution of the force objective $f_\text{F}$ becomes effectively negligible, rendering the target functions nearly identical, which can lead to coincident solutions.

    \begin{table}
            \centering
            \begin{tabular}{l | c | c| c| c|}
                Configuration label                     & H1                 & H1 + finite $F_*$    &  H2 & H3  \\
                Total number of coils                   & $1999$            & $1196$        & $1576$        & $1254$ \\
                Dalí clock coils                             & $916$             & $1027$        & $1397$        & $977$ \\
                Train track coils                            & $183$             & $169$         & $179$             & $277$\\ 
            \end{tabular}
            \caption{Total number of automatically optimised banana coils for the H1, H2, H3 equilibria without a force threshold, and the H1 equilibrium including a finite force threshold, that is a non-zero $F_*$. The coils are categorised as Dalí clock or train-track based on their net twist angle (equation \eqref{eq:twist}.}
            \label{tab:fluxnumbers}
        \end{table}

    \subsection{Optimal banana coil geometries}
\begin{figure}
            \centering
            \begin{subfigure}[t]{0.4\linewidth}
                \centering
                \includegraphics[width=\linewidth]{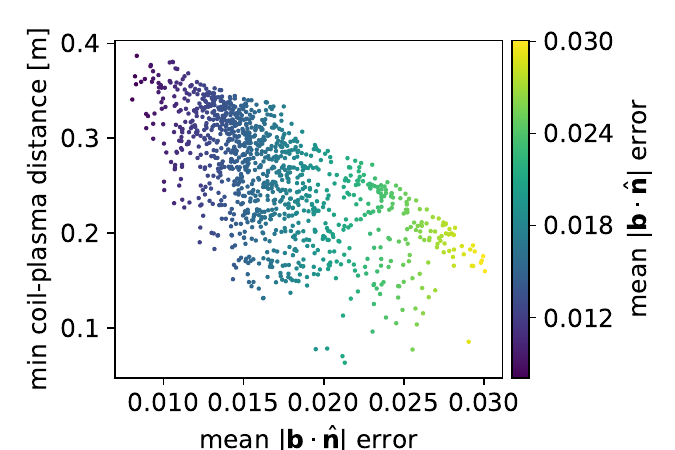}
                \caption{Minimum coil-plasma distance}
                \label{fig:csdist0947}
            \end{subfigure}
            \hspace{0.5cm}
            \begin{subfigure}[t]{0.4\linewidth}
                \centering
                \includegraphics[width=\linewidth]{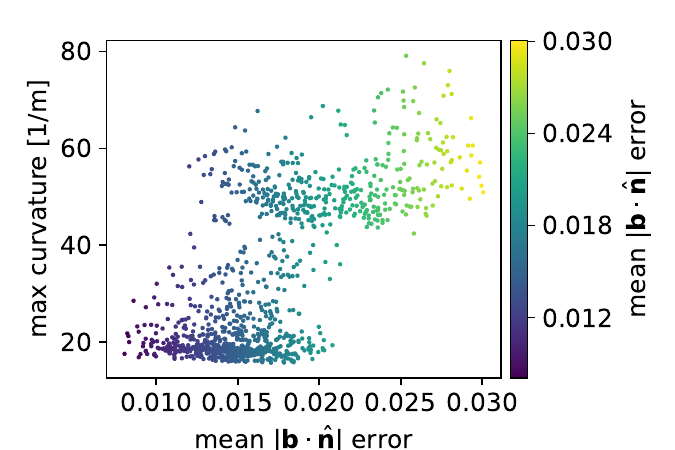}
                \caption{Maximum curvature}
                \label{fig:curvature0947}
            \end{subfigure}
            \caption{The maximum local curvature and minimum coil-plasma separation of the optimised banana coils against the mean normal field error. The optimisations were performed for the H1 Hybrid.
            The colour corresponds to the mean $|\mathbf{b}\cdot \hat{\mathbf{n}}|$ error on the $x$ axis for easier reading of the plots.
            There appear two groups of banana coils characterised by lower and higher maximum curvature values, respectively.}
        \end{figure}

\begin{figure}
            \centering
            \includegraphics[width=0.7\linewidth]{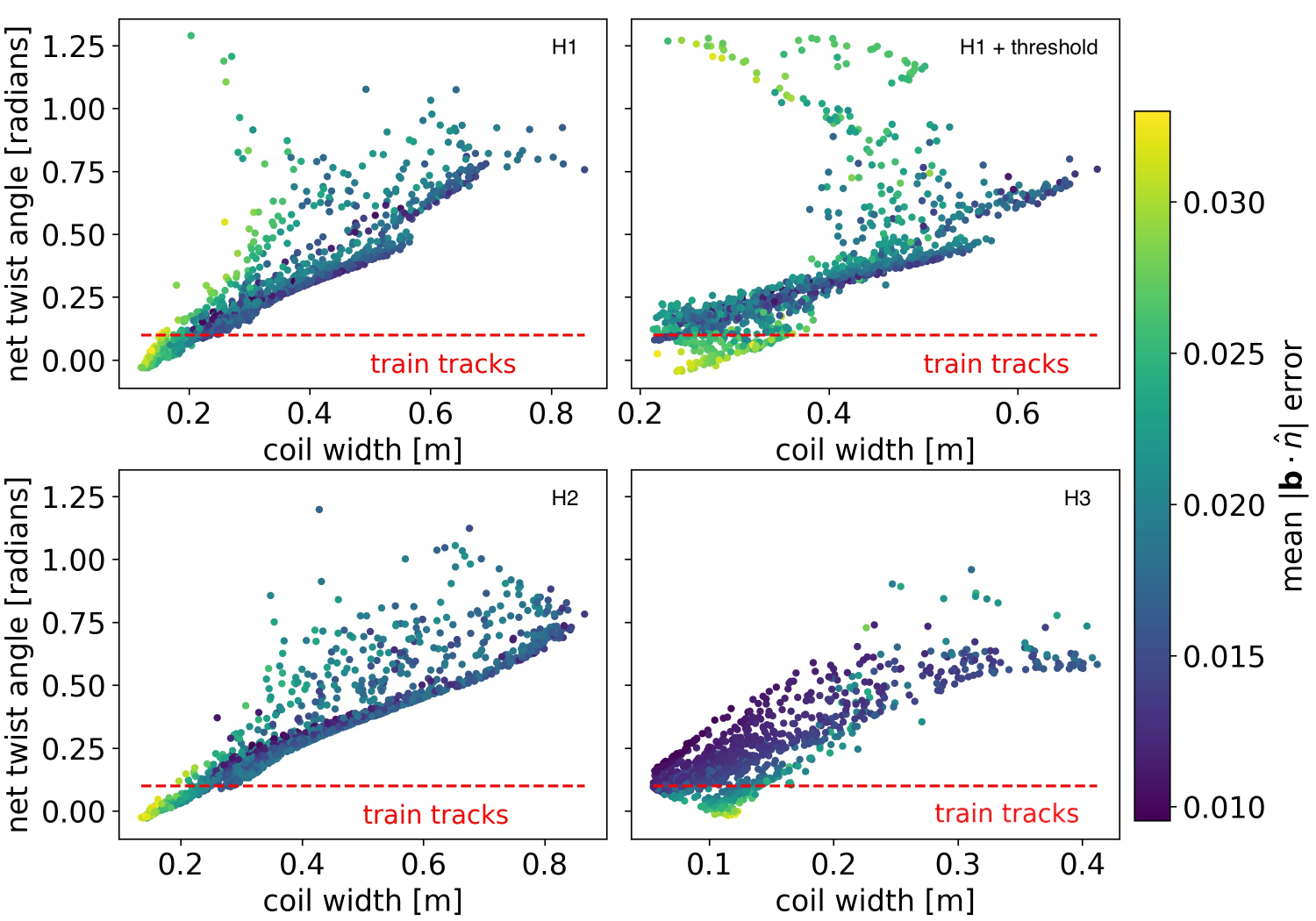}
            \caption{Relationship between the coil width and the net twist angle. The colour scale represents the mean normal field error. Two families of coil solutions are defined: coils with central twist angles below $0.1$ radians are classified as train track coils, while those with larger twist angles are referred to as Dalí clock coils.}
            \label{fig:widthslopediff}
        \end{figure}

    Figure \ref{fig:csdist0947} indicates that increasing the minimum distance between the optimised banana coils and the plasma generally reduces the normal field error. 
    Moreover, we observe a divide in the geometries of force-optimised banana coils.  
    In figure \ref{fig:curvature0947}, the banana coils seem to cluster into two families, characterised by relatively low and relatively high maximum local curvature, respectively.  
    Additionally, we find a strong correlation between the central coil width and the central twist angle, as shown in figure \ref{fig:widthslopediff}.
    Here, thin coils often have two parallel-running filaments that are aligned almost perpendicular to the plasma boundary. Their twist angle remains small, and can even be slightly negative, due to the poloidal curvature of the surface not accounted for in this measure. As a result of this narrow coil shape, the endpoints also exhibit a high curvature.
    Wide coils, on the contrary, tend to have a distinctive twist at the centre. The filament closer to the plasma is generally straighter and follows the shape of a groove, while the outer filament draws out a wave-form, so that the section near the two endpoints aligns with its flat side tangent to the surface. The direction of this twist is in all cases the same for the wide coils, and the endpoints bend at smaller curvatures in comparison to the narrow coils. 
    The field error is noticeably larger for the thin, parallel coils than for the wide, twisted coils.
    We observe a similar separation in the other four optimisations with subtle differences.
    In particular, for the optimisations that included a force threshold, the low-twist banana coil solutions seem to shift to larger widths.
    For the $n_\text{fp}=6$ configuration H3 (bottom right in figure \ref{fig:widthslopediff}), the twisted coil solutions have very small central widths. When inspecting individual coils, it becomes apparent that these coils are most narrow at the centre and widen over the length of the coil. The central coil width in these cases is not representative for the overall coil width. As a consequence, we can observe a shift in the data to smaller widths than an average value would be. 
        
\begin{figure}
            \centering
            \begin{subfigure}[t]{0.32\linewidth}
                \centering
                \includegraphics[width=\linewidth]{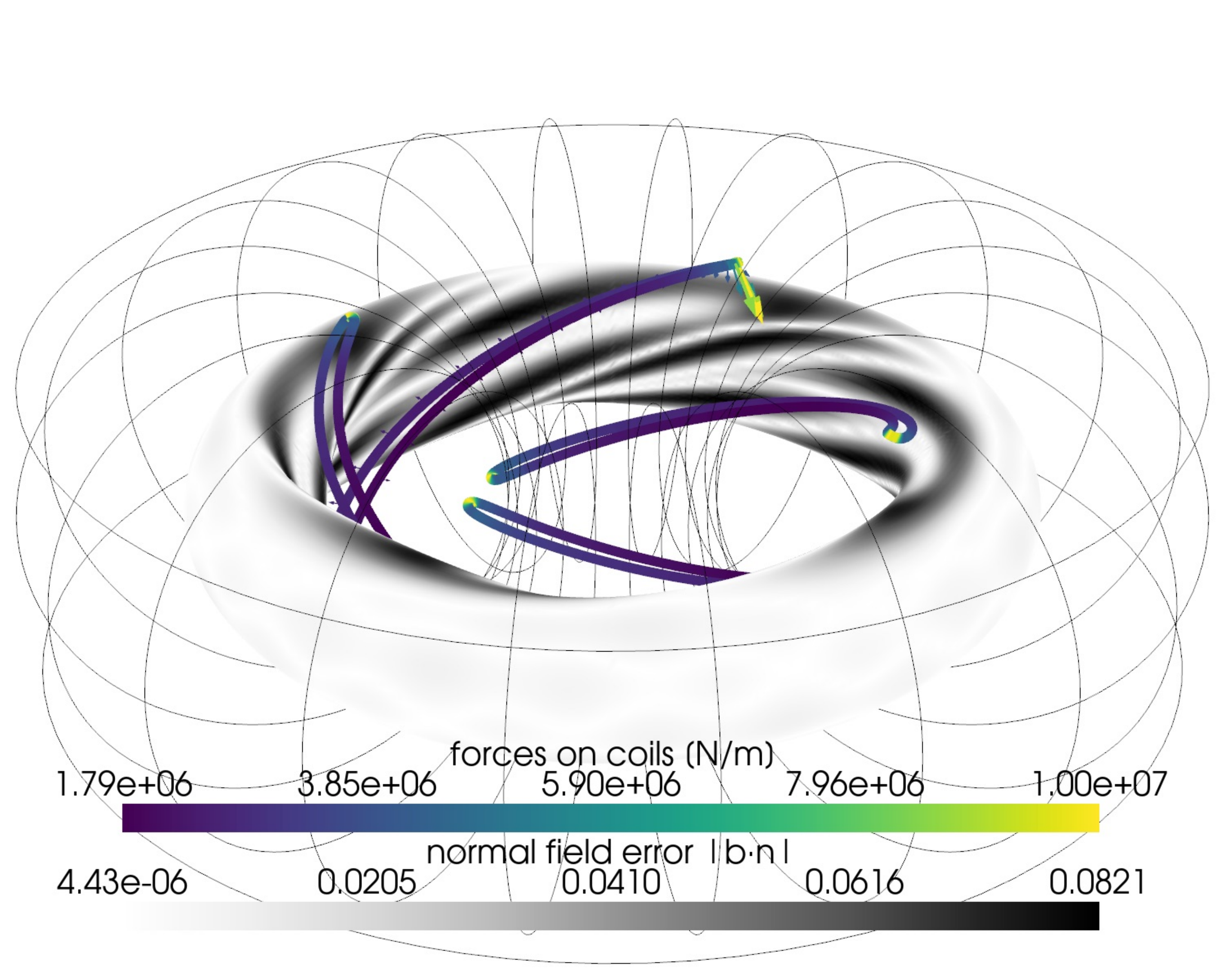}
                \caption{Train track coil}
                \label{fig:traintrack}
            \end{subfigure}
            \begin{subfigure}[t]{0.32\linewidth}
                \centering
                \includegraphics[width=\linewidth]{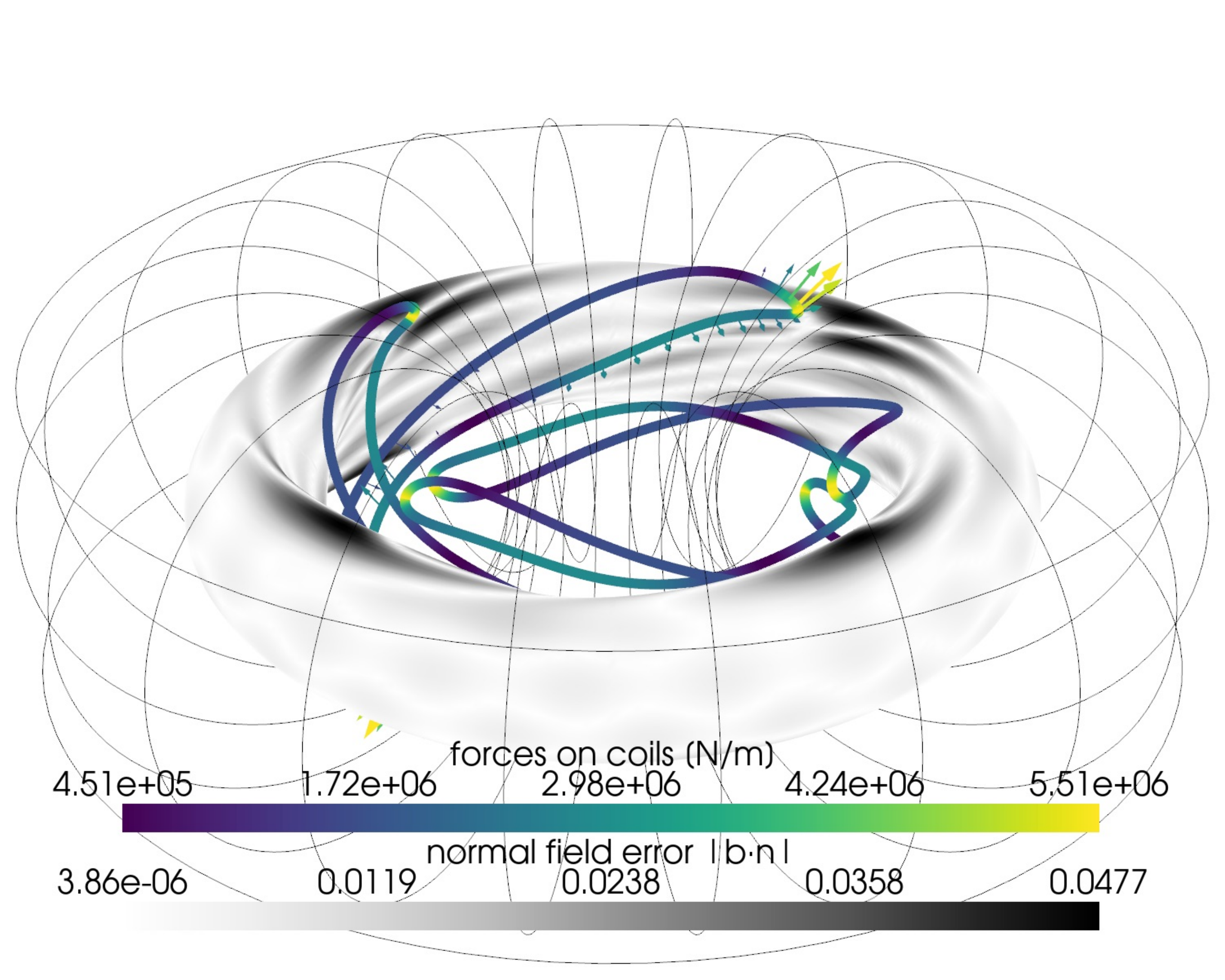}
                \caption{Dalí clock coil}
                \label{fig:daliclock}
            \end{subfigure}
            \begin{subfigure}[t]{0.32\linewidth}
                \centering
                \includegraphics[width=\linewidth]{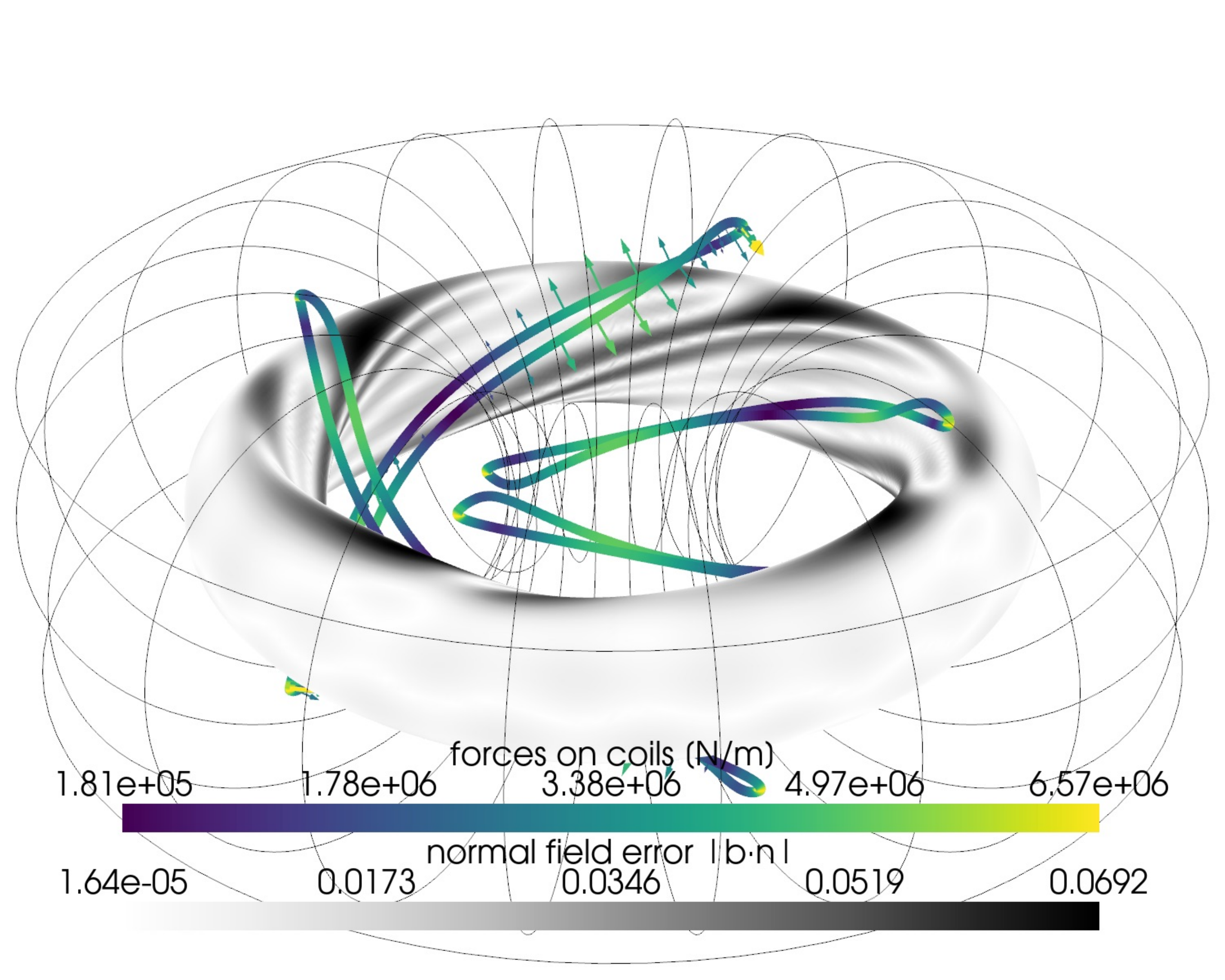}
                \caption{Intermediate coil}
                \label{fig:intermediatecoil}
            \end{subfigure}
            \caption{
            Examples coils from the automated optimisation for the H1 Hybrid equilibrium.
            Panel (a) shows train-track coils, panel (b) Dalí clock coils, and panel (c) an intermediate configuration, in this case obtained using a threshold during the optimisation. The greyscale shading on the flux surface represents the normal field error produced by each coil configuration. The colour scale on the coils indicates the forces experienced by the coils.}
            \label{fig:exampledalitrains}
        \end{figure}

    Based on this characteristic difference in the geometry, we divide banana coils into two categories. We name the two groups according to their visual appearance.
    The narrow parallel coils, as those in figure \ref{fig:traintrack}, are referred to as \textit{train track }coils \footnote{One can picture a train running up the two parallel filaments as if they were rails.}, and the twisted wide coils, as those in figure \ref{fig:daliclock}, are called \textit{Dalí clock} coils \footnote{There is a resemblance to the melting clocks in Salvador Dalí's ``The Persistence of Memory'': their flat sides begin to wrap around the surface, as though they were slowly melting onto it.}.
    When looking through the outline of a coil as if it was a window frame, one could see the plasma boundary with the wide, twisted Dalí clock coils, but not for the thin, parallel train track coils.
    The two families of coils are continuously connected in the coil parameter space, and specific boundary values chosen to distinguish them are somewhat arbitrary. While some coils clearly belong to one or another family, there are also some intermediate coils like that shown in figure \ref{fig:intermediatecoil}.  To get an insight into the qualitative differences between the two families of coil solutions, we decided on defining train track coils to be those with a central twist angles below $0.1$ radians, and all other coils to be the Dalí clock coils. This threshold value is a crude estimate, and as apparent from figure \ref{fig:widthslopediff}, wrongly classifies a few coils. Table \ref{tab:dalitrains} gives an overview of the properties of the coils shown in \ref{fig:exampledalitrains}, and table \ref{tab:fluxnumbers} summarises the number of train track and Dalí clock coils found for each configuration.

        \begin{table}
            \centering
            \begin{tabular}{l | c| c| c| }
                 coil type                           &  Dalí clock                            & Intermediate                      & Train track       \\[3pt]
                 $|\mathbf{b}\cdot \hat{\mathbf{n}}|_{\max}$ &  $4.71\%$                    & $6.90\%$                          &  $8.21\%$ \\
                 $|\mathbf{b}\cdot \hat{\mathbf{n}}|_\text{mean}$ & $1.07\%$                & $1.86\%$                          &  $2.42\%$  \\[3pt]
                 curve length [$\mathrm{m}$]                      &   $12.07$                   & $10.85$           & $10.51$ \\
                 curvature [$\mathrm{m}^{-1}$]                          &   $14.77$             & $55.65$      & $61.12$ \\
                 net twist angle                    &   $0.5083$                            &   $0.1166$                        & $-0.0157$ \\
                 coil current [$\mathrm{MA}$]                       &  $1.14$                 & $2.74$              & $2.46$ \\[3pt]
                 integrated force [$\mathrm{N}$]         &   $2.38\times10^7$       & $3.31\times10^7$    &  $2.88\times10^7$\\
                 max. force [$\mathrm{ N/m}$]            &   $5.51\times10^6$       & $6.58\times10^6 $ &  $1.00\times10^7$\\
                 max. TF force [$\mathrm{ N/m}$]         &   $1.09\times10^6 $      & $1.03\times10^7 $  &  $1.08\times10^7 $ \\
                 max. self force [$\mathrm{ N/m}$]       &    $1.25\times10^6 $     & $8.99\times10^6 $ &  $8.01\times10^6 $\\
                 max. mutual force  [$\mathrm{ N/m}$]    &    $2.79\times10^5 $     & $2.77\times10^5 $ &  $1.59\times10^5 $\\
            \end{tabular}
            \caption{Table comparing the properties of the three banana coils shown in figure \ref{fig:exampledalitrains}.}
            \label{tab:dalitrains}
        \end{table}

    \subsection{Coil-plasma distance}
\begin{figure}
            \centering
            \includegraphics[width=0.5\linewidth]{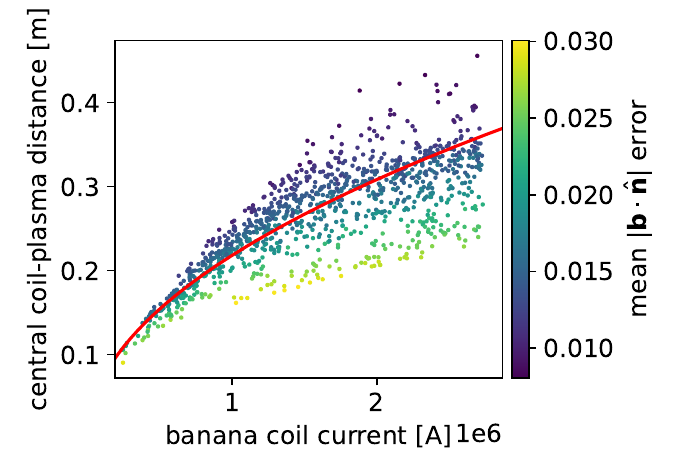}
            \caption{Central curve–plasma distance as a function of the banana coil current. The colour scale represents the mean field error, $|\mathbf{b} \cdot \hat{\mathbf{n}}|$. The data is fitted by the relation $I = d^2 / (2.10 \times 10^{7})$, shown as a red line.}
            \label{fig:currentcenterdist2}
        \end{figure}

    In an optimisation, the coil position relative to the plasma is adjusted to best reproduce the target magnetic field at the boundary. Changing the coil current $I$ scales the magnetic field strength linearly, and thereby influences the optimal coil to plasma separation. 
    For the optimised banana coils, this current dependence of the coil-plasma separation is shown in figure \ref{fig:currentcenterdist2}.
    Note that despite optimising coils for the same target field, their realised boundary field always deviates slightly from the target. As a result, at any point on the boundary, there is a spread in the achieved magnetic field strengths generated by different coils, which manifests as a range of different normal field errors and a variation in the curve-plasma distances.
    In the coil model of two-straight wires from equation \eqref{eq:bfieldtwostraight}, when sufficiently far, coils that achieve the same magnetic field strength follow a relationship of the form $r \sim \sqrt{I}$, where $r$ here takes on the role of the curve-plasma distance.
    We fit a curve of this form, the red line in figure \ref{fig:currentcenterdist2}, which approximately follows the contour of constant mean field error on the entire boundary, $|\mathbf{b}\cdot \hat{\mathbf{n}}|_\text{mean}$.

\begin{figure}
            \centering 
        \begin{subfigure}[t]{0.45\linewidth}
            \centering
            \includegraphics[height=4.2cm]{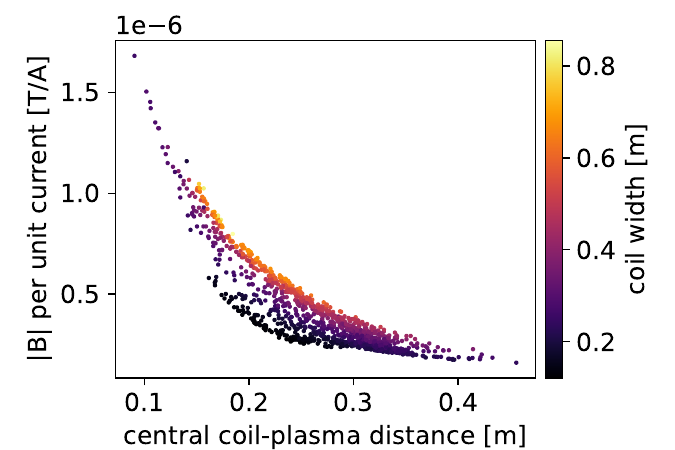}
            \caption{Comparison with coil width. Here, the colour scale indicates the central coil width. The coils with the largest curve–surface distance have widths of approximately $0.3 \mathrm{m}$.}
            \label{fig:BCSdist}
        \end{subfigure}
        \hspace{0.5cm}
        \begin{subfigure}[t]{0.45\linewidth}
            \centering
            \includegraphics[height=4.2cm]{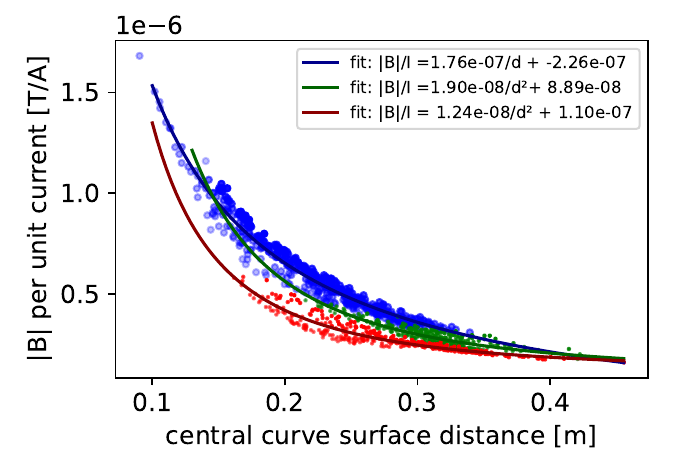}
            \caption{Trendlines for coils with $d<a=w/2$ are shown in blue. The remaining coils satisfy $d \gtrsim a$ and are further categorised by width: green corresponds to $w > 0.3,\mathrm{m}$ and red to $w < 0.3,\mathrm{m}$. The opacity indicates coil width, with more transparent points representing smaller widths and more opaque points representing larger widths.}
            \label{fig:BCSdist_trends}
        \end{subfigure}
        \caption{Magnetic field strength per unit current near the plasma centre from a single banana coil, $|\bfB_\text{b}|/I$, against the central coil–surface distance for the coils from the automated optimisation for the first equilibrium. Figure (a) coil width dependence. Figure (b): numerical fit.}
        \end{figure}

    Computing the ratio of the magnetic field strength to the coil current $|\bfB|/I$ reveals the underlying geometric dependence of the coil–plasma separation.
    Figure \ref{fig:BCSdist} shows the relation between the central curve-plasma distance and the magnetic field $\bfB_\text{b}$ generated by individual banana coils and evaluated at the central surface point $R_s$, per unit current, i.e. $|\bfB_\text{b}|/I$ \;\footnote{To avoid confusion, note that most of the magnetic field due to one banana coil at the central surface point on boundary, $\bfB_\text{b}$, arises from the component tangent to the plasma boundary, such that $|\bfB_\text{b}| \simeq |\bfB_\text{b} \times \hat{\mathbf{n}}|$.}.
    Again, we find that the optimised coils tend to fall into two groups according to how rapidly $|\bfB_\text{b}|/I$ decreases with increasing distance. The magnetic field strength of narrow coils diminishes more quickly and at shorter distances compared to wide coils. This behaviour is consistent with the analytical model of two parallel straight wires: as the two filaments get closer together, the magnetic fields from the opposing currents increasingly cancel, reducing the net field strength. 
    Figure \ref{fig:BCSdist} illustrates the transition from narrow coils that move away from the plasma, and thereby increase the coil–plasma separation, to wide coils that, beyond a critical width, begin to shift back toward the boundary as the inner filament again approaches the plasma. Earlier, we found that the narrow coil solutions correspond to the train track coils, and the wide coils to Dalí clock coils (figure \ref{fig:widthslopediff}).
    Informed by the geometry, we found that narrow, far away coils, which classify as train track coils, follow a parallel double-wire model description, where $|\bfB|/I \sim 1/r^2$ from equation \eqref{eq:bfieldtwostraight}, as shown with the red trendline in figure \ref{fig:BCSdist_trends}. 
    The wide coils, which are primarily Dalí clock coils, here seem to transition between a double-wire to a single-wire description from equation \eqref{eq:bfieldonestraight}, that is $|\bfB|/I \sim 1/r$, as the coil-plasma distance becomes small compared to the width ($d\lesssim a=w/2$). 
    These fits include an additional constant to account for coil contributions beyond these simple models, such as curved features or the respective coil width. A smaller added coefficient for the far away, Dalí clock coils suggests a weaker dependence on the coil width than for the narrow train track coils.

    \subsection{Trade-off between force and normal field error}
\begin{figure}
            \centering
            \includegraphics[width=0.5\linewidth]{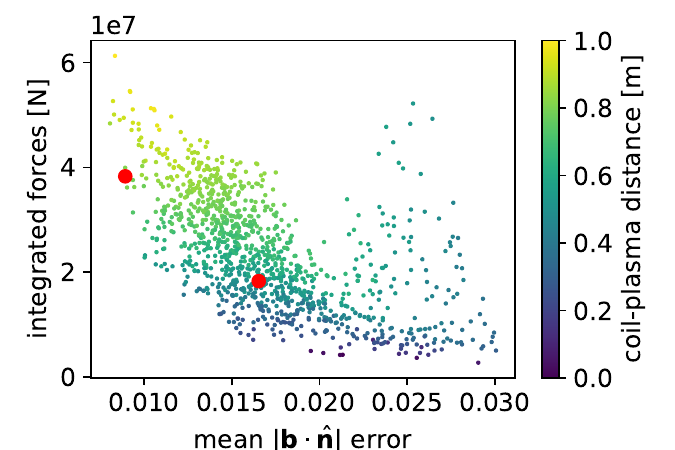}
            \caption{Trade-off between the integrated force magnitude, $\int_\mathcal{C} \left| \mathrm{d}\mathbf{F} / \mathrm{d}\mathbf{l} \right| \mathrm{d}l$, and the mean normal field error, $|\mathbf{b} \cdot \hat{\mathbf{n}}|$, for a zero threshold force optimisation. The colour scale represents the coil-surface distance. Coil solutions located toward the lower-left boundary of the plot, corresponding to both low integrated forces and low normal field error, represent the best-performing configurations with respect to these two objectives. The red dots correspond to the train track and Dalí coils in figure \ref{fig:exampledalitrains}.}
            \label{fig:force-BdotN}
        \end{figure}

    In the coil optimisations, both the integrated magnitude of the forces, $\int_\mathcal{C}|\text{d}\mathbf{F}/\text{d}\mathbf{l}|\text{d}l$, and the normal field error are penalised. 
    Figure \ref{fig:force-BdotN} shows how the integrated forces and the mean normal field error compare for different coil solutions. In the metric of these two objectives, banana coils with the least forces and lowest normal field error, found in the bottom left region of the plot, are the best-achieving ones. We see that lower forces correspond to larger field errors. Reducing the forces on the banana coils therefore comes at the expense of how accurately they can reproduce the equilibrium. 
    Consistent with the findings of \cite{hurwitz_electromagnetic_2025}, we observe a positive correlation between the coil-plasma distance and the forces: just as for the modular coils, banana coils further away from the plasma necessarily correspond to radially more inward coil positions, exposing them to larger toroidal field strengths. In this case, the banana coils experience an additional effect from their flexibility in the choice of coil current. The coil-plasma distance of optimised coils increases with coil current, and thereby also the pointwise forces.
    
    \subsection{Current scaling and upper bound on forces}
\begin{figure}  
            \centering
            \begin{subfigure}[t]{0.45\linewidth}
                \centering
                \includegraphics[width=\linewidth]{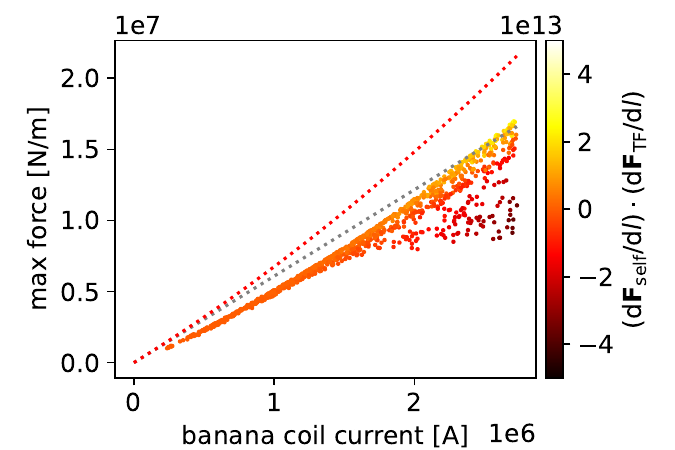}
                \caption{Optimisation with zero threshold.}
                \label{fig:forcecurrent_zero}
            \end{subfigure}
            \hspace{0.5cm}
            \begin{subfigure}[t]{0.45\linewidth}
                \centering
                \includegraphics[width=\linewidth]{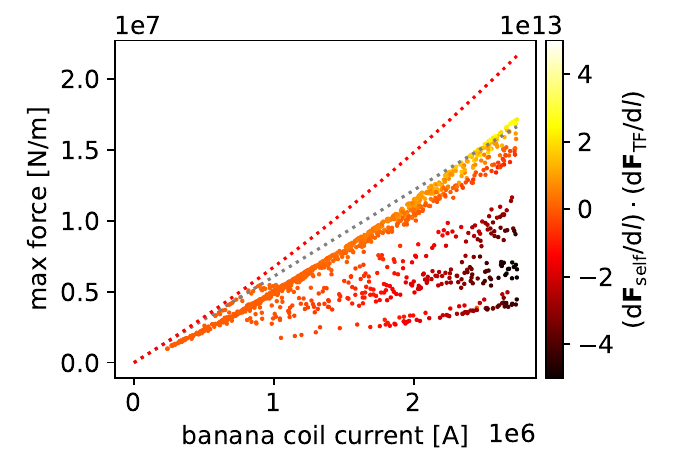}
                \caption{Optimisation with a finite threshold.}
                \label{fig:forcecurrent_finite}
            \end{subfigure}
            \caption{Integrated force on the banana coil as a function of the maximum force for optimisations performed with (a) zero threshold and (b) a finite threshold. The colour scale represents the alignment between the self and toroidal magnetic fields, $\bfB_\text{self}\cdot \bfB_\text{TF}$. The gray curve indicates the force bound due to the toroidal field alone, while the red curve shows the bound including toroidal, self, and mutual-field contributions, which is estimated using equation \eqref{eq:forcebounds} ($R_{\min}=1.5\textrm{m}$ and $d_{\min}=0.3\textrm{m})$.}
            \label{fig:forcecurrent}
        \end{figure}

    We estimated an upper bound for the magnetic forces acting on banana coils using equation \eqref{eq:forcebounds}, and found the optimised coils to lie below, as shown in figure \ref{fig:forcecurrent}. In fact, most coil solutions adhere to an even lower limiting value and only individual outliers at low currents with unfavourable geometries have larger forces -- still below the estimated bound -- that can arise for example from particularly small spacings between filaments, i.e. smaller $d_{\min}$.
    For banana coil solutions below this upper front, the self and toroidal field forces partially cancel. 
    The alignment between these two forces is indicated by the scalar product of the respective magnetic field contributions, $\bfB_\text{self}\cdot \bfB_\text{TF}$.
    When this scalar product is zero, $\bfB_\text{self}$ and $\bfB_\text{TF}$ are perpendicular, for positive values, the two fields are parallel and the forces add, and conversely, for negative values, the fields are anti-parallel, and the forces subtract.
    The coils that experience the highest forces are those with large currents. At the location of the maximum force along each coil, the corresponding values of $\bfB_\text{self}\cdot \bfB_\text{TF}$ are found to be large and positive, meaning that the self and toroidal fields unfavourably align for these coils. At these large currents, the self-forces become increasingly significant and comparable in size to the toroidal forces, contributing to the increased scalar product, especially in comparison to coils with less current.
    The further a banana coil is below this upper force bound, the more the self and toroidal fields counterbalance.
    We found significantly more coil solutions that reduce the maximum forces by this field alignment mechanism for the optimisations using a finite force threshold value compared to those without, as seen by comparing figures \ref{fig:forcecurrent_zero} and \ref{fig:forcecurrent_finite}.

    \subsection{Toroidal field forces}
\begin{figure}
            \centering
            \begin{subfigure}[t]{0.45\linewidth}
                \centering
                \includegraphics[width=\linewidth]{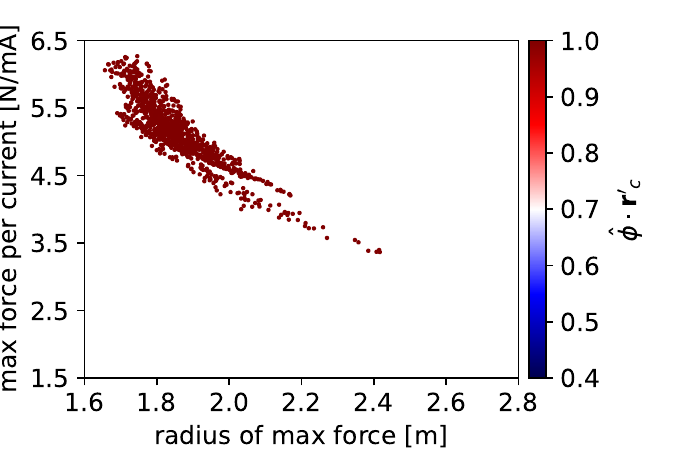}
                \caption{Optimisation with zero threshold.}
            \end{subfigure}
            \hspace{0.5cm}
            \begin{subfigure}[t]{0.45\linewidth}
                \centering
                \includegraphics[width=\linewidth]{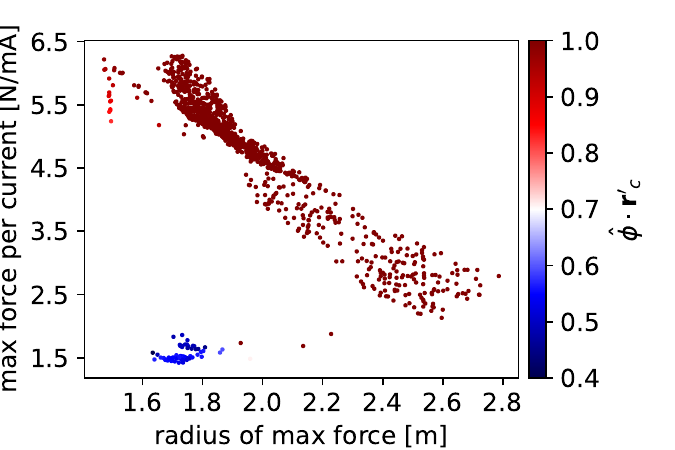}
                \caption{Optimisation with finite threshold.}
            \end{subfigure}
            \caption{Maximum force on the banana coil per unit current versus the radius at which the maximum force occurs, for cases with (a) zero threshold and (b) finite threshold. The colour scale represents the magnitude of the cross product between the coil tangent vector and the toroidal unit vector, $\bfr_c' \times \hat{\boldsymbol{\phi}}$, at the location of maximum force along a coil.}
            \label{fig:forceradius}
        \end{figure}

    Banana coils can reduce the forces from the toroidal field either by positioning to larger radial locations, where the toroidal magnetic field is weaker, or by orienting the coil tangent vector, $\bfr_c'$, to be aligned with the toroidal field, $\bfB_\text{TF}$, in the $\hat{\boldsymbol{\phi}}$ direction. When the cross product $\bfr_c' \times \hat{\boldsymbol{\phi}}$ is zero, the tangent vector is parallel to the toroidal field, and the coil section experiences no force from it. 
    Figure \ref{fig:forceradius} illustrates how the first of these mechanisms lowers the force on the coils: the maximum force per unit current decreases as the radial location at which it occurs increases. We here normalised the pointwise force by dividing by the banana coil current, such that the current scaling of the forces is factored out.
    Only in the optimisation with a finite force threshold, another group of coil solutions appear, which make use of the second mechanism and have a low $\bfr_c' \times \hat{\mathbf{\phi}}$ at the location of their maximum force. Among these particular coil solutions is the coil shown in figure \ref{fig:intermediatecoil}. For this case, we found the largest forces located along the long side of the coil filament further away from the plasma. In this straight section, the angle to the toroidal field can be significantly shallower than at the endpoints, where the coil curve necessarily makes a right angle to the toroidal field. The forces at the endpoints of this coil are reduced both by an increased radial location, and by largely cancelling the toroidal with the self forces.

\begin{figure}  
            \centering
            \begin{subfigure}[t]{0.45\linewidth}
                \centering
                \includegraphics[width=\linewidth]{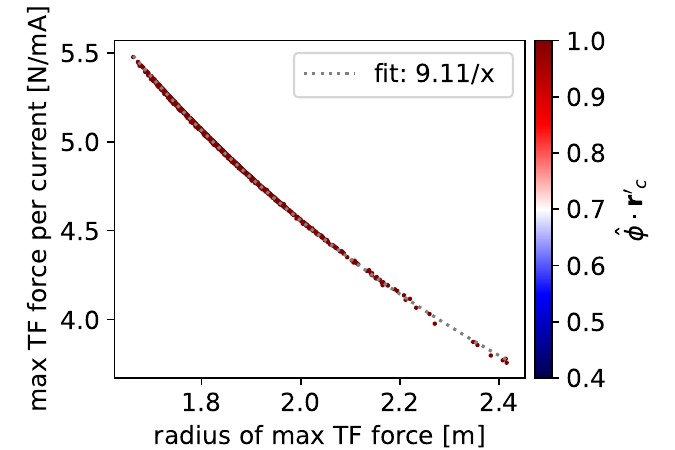}
                \caption{Optimisation with zero threshold.}
                \label{fig:forceradiusTF_zero}
            \end{subfigure}
            \hspace{0.5cm}
            \begin{subfigure}[t]{0.45\linewidth}
                \centering
                \includegraphics[width=\linewidth]{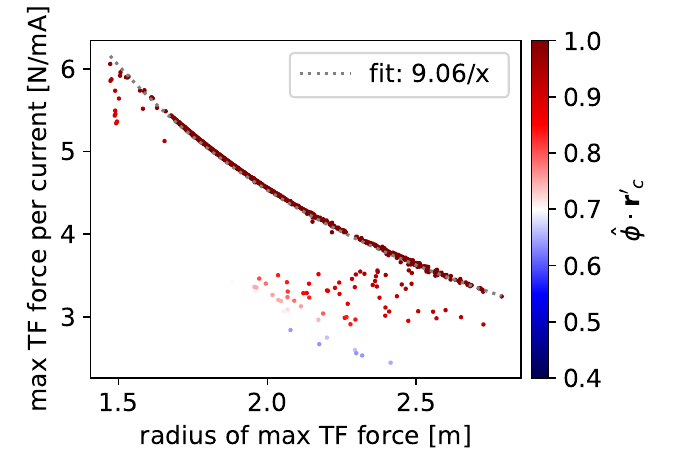}
                \caption{Optimisation with a finite threshold.}
                \label{fig:forceradiusTF_threshold}
            \end{subfigure}
            \caption{Maximum toroidal field force on the banana coil per unit current versus the radius at which this maximum TF force occurs, for cases with (a) zero threshold and (b) finite threshold. Thecolour scale represents the magnitude of the cross product between the coil tangent vector and the toroidal unit vector, $\bfr_c' \times \hat{\mathbf{\phi}}$, at the location of the maximum TF force. A $1/R$ curve is included, indicating the radial dependence of the maximum TF force per unit banana coil current.}
            \label{fig:forceradiusTF}
        \end{figure}

    To study the second mechanism in more detail, we found the location along each coil, where the TF force is largest. For most coils, this location coincides with the turning point, where $\bfr_c' \times \hat{\boldsymbol{\phi}} \approx 1$. However, also other coil sections can be oriented perpendicular to the toroidal field, for example at the twist in the filament further away from the plasma in a Dalí clock coil. 
    Since the toroidal field falls off as $1/R$ for radial locations $R$, the ratio of the maximum TF force per banana coil current should also follow this radial dependence. Figure \ref{fig:forceradiusTF} includes a trendline with this relationship, which agrees well with the optimised coil solutions. 
    Coils with values below this curve increasingly align the tangent vector with the toroidal direction at the locations where they experience the largest forces due to the TF field.

    \subsection{Force distribution along banana coils}
    Comparing the three coils in figure \ref{fig:exampledalitrains} -- a train track, a Dalí clock, and an intermediate coil -- reveals differences in the spatial distribution of forces along the filament. The train-track coils exhibit highly localised forces near the endpoints, but low forces over the majority of their length. By contrast, the Dalí-clock coils experience appreciable forces on coil section farther away from the plasma. Because these different distributions can yield comparable integrated forces, the associated trends are not well-reflected in any of the preceding plots. This disparity nevertheless indicates that the corresponding local minima of the coil families in the optimisation landscape may favour certain mechanisms of force reduction.

\section{Discussion and Conclusions} \label{sec:conclusion}

    Across the three different equilibria, we observed similar behaviour in the optimisation of banana coils.  
    It should be noted here that the analysed data set is arbitrarily sampled; although it reflects the optimisation behaviour, it does not necessarily represent the entire optimisation landscape. 
    We identified two coil families associated with different geometric characteristics, which we refer to as train track and Dalí clock coils, and which may indicate two distinct local minima. These two groups are not strictly separate, and intermediate coil shapes exist. They differ in coil width and twist angle, which in turn shape the structure of the magnetic field they produce. The observed division of optimised coils into two groups with different maximum curvatures further supports the presence of these geometric distinctions. 
    The measures that we used to characterize the coils are somewhat limited in making this distinction, and do not capture some of the relevant coil features. Especially for the H3 boundary ($n_\text{fp}=6$, $\iota_\text{ext}=0.05$), we found these quantities to fail in this regard, for example in neglecting the variation of the coil width. While they have been effective to identify trends, different geometric measures may be necessary to fully describe the important characteristics in the shape of banana coils. A different representation of the coils than using Fourier coefficients may also ease the search for adequate parameters, while likely also simplifying the banana coil optimisation landscape.
    
    The proposed simple coil models based on either two parallel straight wires or a single filament can describe key features in the magnetic field of the two coil families, and thereby provide an understanding on how the optimal coil–plasma distance increases as the coil current is raised. In particular, our results illustrate how the coil-plasma distance can be controlled by choosing the current, which then again is interrelated to other coil properties:
    in optimised configurations, higher coil currents can increase the separation between the banana coil and the plasma boundary, which in turn can reduce the field error. 
    This observed improvement is likely linked to reduced coil ripple when the coils are placed farther away. 
    At the same time, the current directly scales the electromagnetic forces, and, as a secondary consequence, the increased coil-plasma distance results in coils that are positioned in regions with larger toroidal magnetic fields, thereby further amplifying the forces acting on the coils.
    We found that these forces can be mitigated by several mechanisms: by cancellation of the self and TF forces, by maximizing the radial location at which a coil is subjected to the highest TF forces, by aligning the coil curve's tangent vector with the toroidal field, and by reducing the current in the coil. 
    Using a force threshold value can help steer these mechanisms, though a more careful analysis of the effect of employing different values should be conducted first to guide an appropriate threshold choice. 
    It may be insightful to study the effect of including other types of electromagnetic force, such as the torque, net force, or shear, on optimised coil solutions to reveal potential trade-offs in the reduction of certain force types. 
    It is important to recognise that, for the coil geometries we have found, the small-parameter assumption used in the regularisation procedure to compute self-forces in \cite{hurwitz_efficient_2024} may no longer be valid, as the radius of curvature can be comparable to the coil thickness. To assess the applicability and accuracy of this method for banana coils, it may therefore be necessary to compare the pointwise forces obtained from the regularised Biot–Savart formulation with those computed using finite-element solvers.

    We have presented results for only a very limited number of targeted Hybrid equilibria, but many more shapes and configurations exist. 
    The optimised coils are particular to these specific boundaries, though one might expect similar trends to appear for other Hybrids.
    The observed differences even between just the three equilibria studied is indicative for how the plasma geometry correlates to coil compatibility, and for how little we understand of this relationship. It remains an open research question which equilibrium features determine how well coils can reproduce them. Ongoing and future work in this area \citep{sengupta2026optical, sengupta2026compact, rodriguez2026estimating, kappel_magnetic_2024} may be particularly useful for this application of banana coils.

\ack{The first author would like to thank Ludovic Rais, Chris Smiet, Alan Goodman, Georgia Acton, and Per Helander for stimulating conversations.}

\begin{figure}
    \centering
    \begin{subfigure}[b]{0.32\linewidth}
        \centering
        \includegraphics[height=3cm]{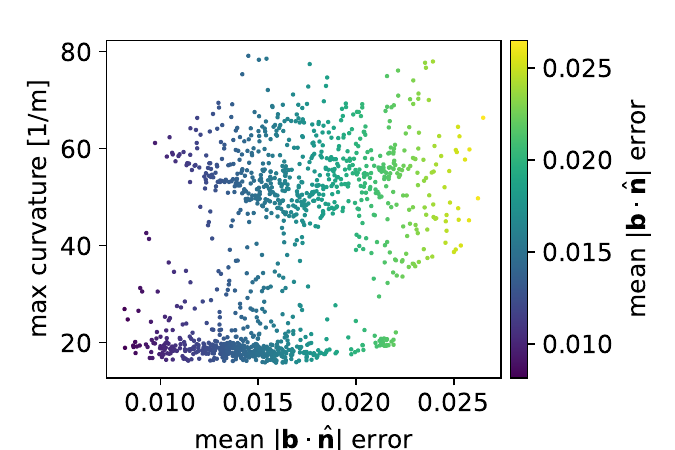}
    \end{subfigure}
    \begin{subfigure}[b]{0.32\linewidth}
        \centering
        \includegraphics[height=3cm]{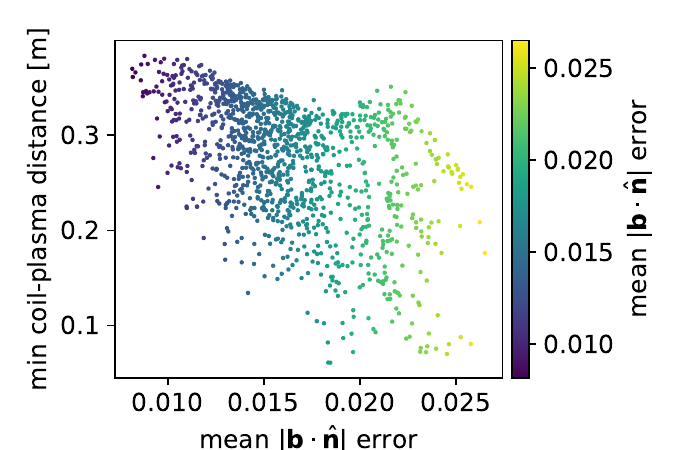}
    \end{subfigure}
    \begin{subfigure}[b]{0.32\linewidth}
        \centering
        \includegraphics[height=3cm]{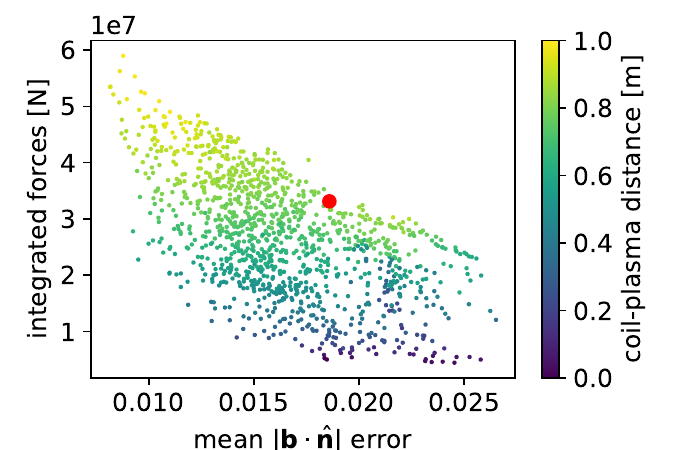}
    \end{subfigure}
    \\
    \begin{subfigure}[b]{0.32\linewidth}
        \centering
        \includegraphics[height=3cm]{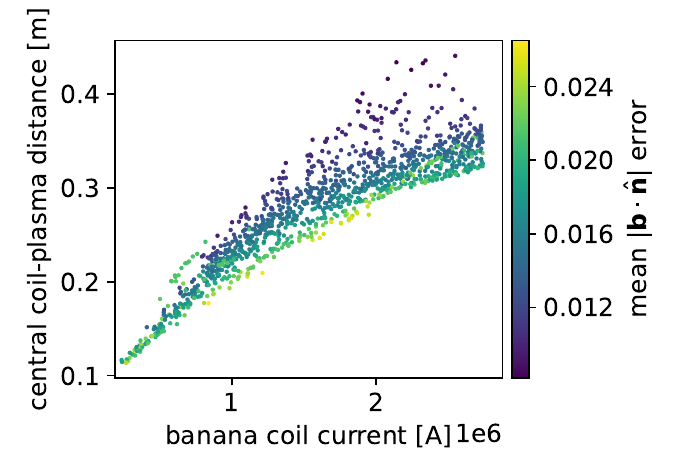}
    \end{subfigure}
    \begin{subfigure}[b]{0.32\linewidth}
        \centering
        \includegraphics[height=3cm]{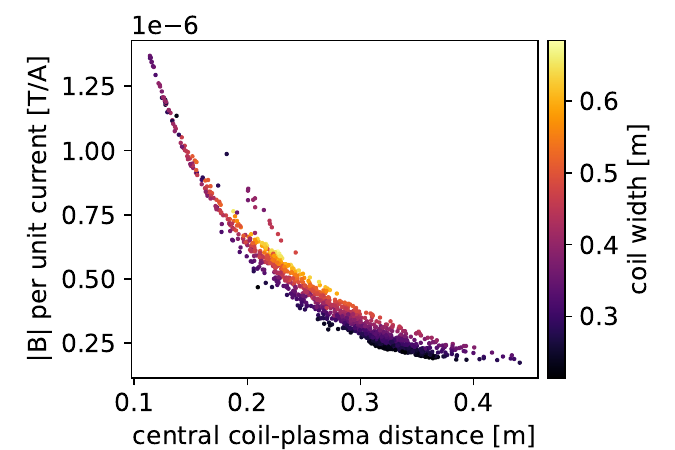}
    \end{subfigure}
    \begin{subfigure}[b]{0.32\linewidth}
        \centering
        \includegraphics[height=3cm]{Plots/1_nfp4_0947/threshold_4e6/0_forcecurrent.pdf}
    \end{subfigure}
    \\
    \begin{subfigure}[b]{0.32\linewidth}
        \centering
        \includegraphics[height=3cm]{Plots/1_nfp4_0947/threshold_4e6/0_forceradius_max.pdf}
    \end{subfigure}
    \begin{subfigure}[b]{0.32\linewidth}
        \centering
        \includegraphics[height=3cm]{Plots/1_nfp4_0947/threshold_4e6/0_forceradius_onlyTF.pdf}
    \end{subfigure}
    \caption{Various plots for optimised coils with a force threshold of $1.75T\times I_\text{banana}$ for the H1 boundary (with $n_\text{fp}=4$, and $\iota_\text{ext} =0.05$). The red dot corresponds to the intermediate coil in figure \ref{fig:exampledalitrains}.}
    \label{fig:data1}
\end{figure}

\begin{figure}
    \centering
    \begin{subfigure}[b]{0.32\linewidth}
        \includegraphics[height=3cm]{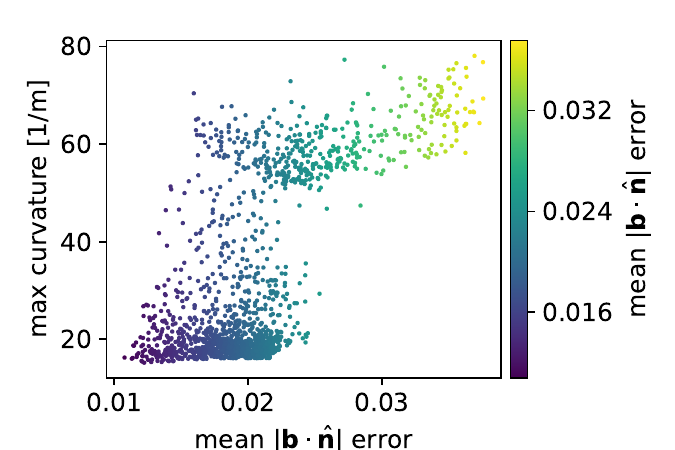}
    \end{subfigure}
    \begin{subfigure}[b]{0.32\linewidth}
        \centering
        \includegraphics[height=3cm]{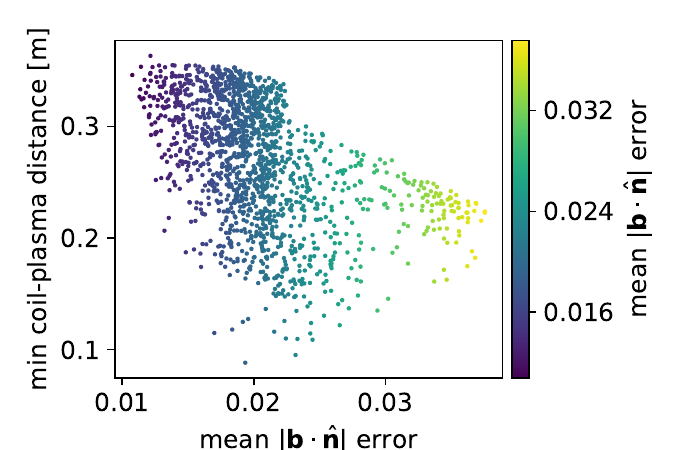}
    \end{subfigure}
    \begin{subfigure}[b]{0.32\linewidth}
        \centering
        \includegraphics[height=3cm]{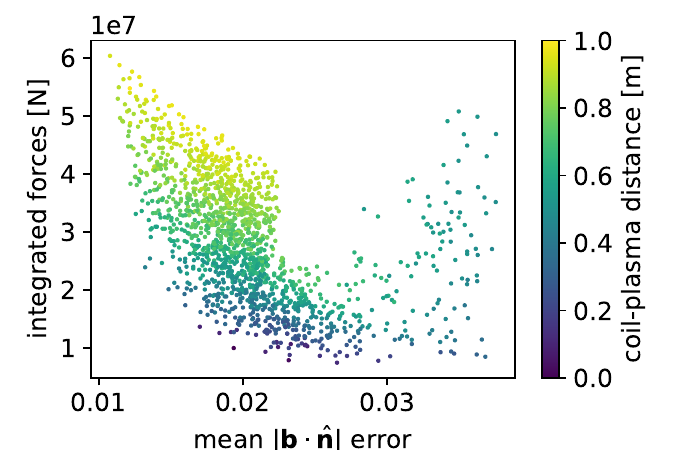}
    \end{subfigure}
    \\
    \begin{subfigure}[b]{0.32\linewidth}
        \centering
        \includegraphics[height=3cm]{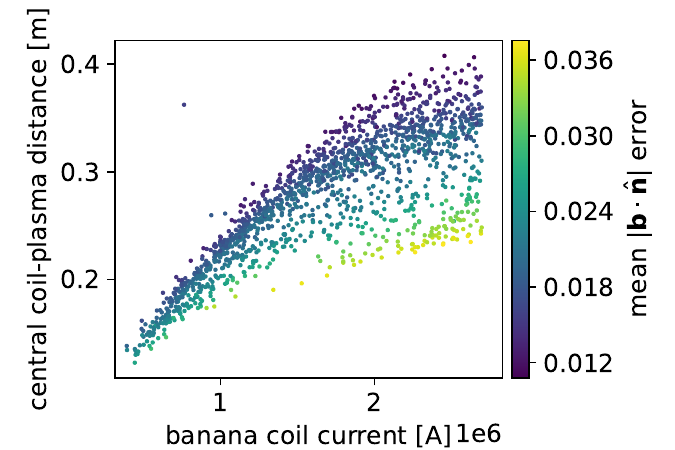}
    \end{subfigure}
    \begin{subfigure}[b]{0.32\linewidth}
        \centering
        \includegraphics[height=3cm]{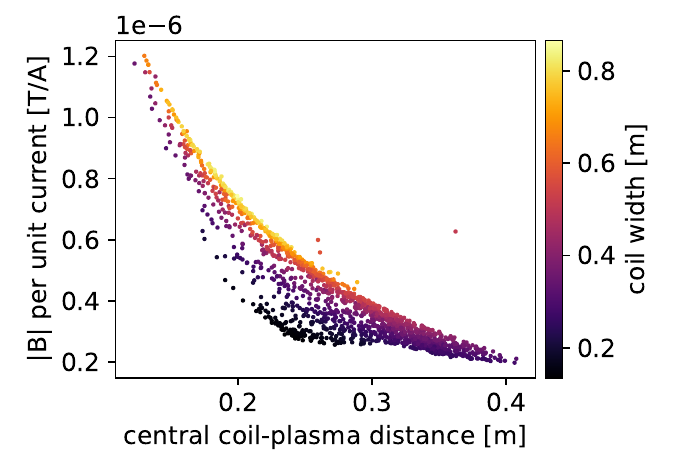}
    \end{subfigure}
    \begin{subfigure}[b]{0.32\linewidth}
        \centering
        \includegraphics[height=3cm]{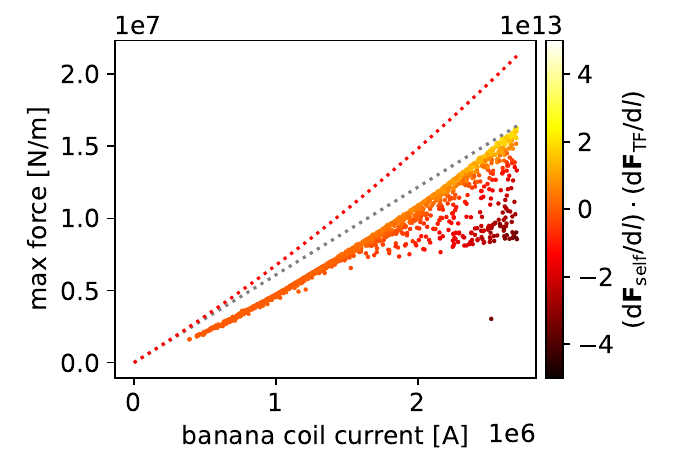}
    \end{subfigure}
    \\
    \begin{subfigure}[b]{0.32\linewidth}
        \centering
        \includegraphics[height=3cm]{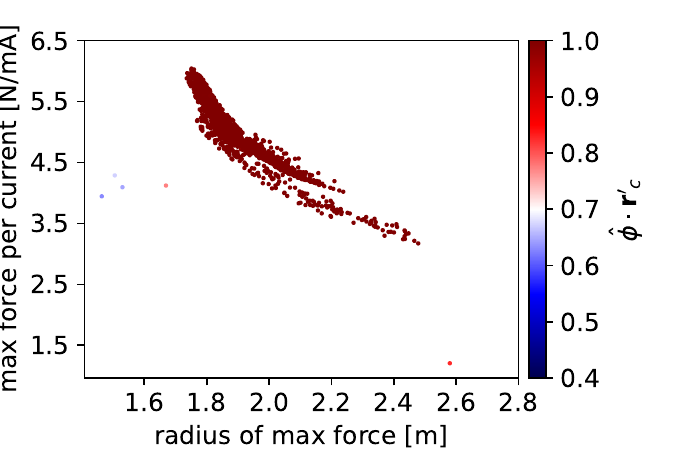}
    \end{subfigure}
    \begin{subfigure}[b]{0.32\linewidth}
        \centering
        \includegraphics[height=3cm]{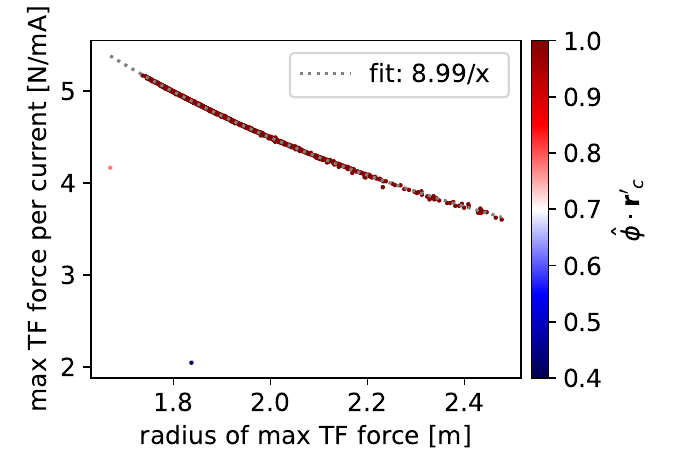}
    \end{subfigure}
    \caption{Various plots for optimised coils with zero force threshold for the H2 boundary(with $n_\text{fp}=4$, and $\iota_\text{ext} =0.15$).}
    \label{fig:data2}
\end{figure}

\begin{figure}
    \centering
    \begin{subfigure}[b]{0.32\linewidth}
        \includegraphics[height=3cm]{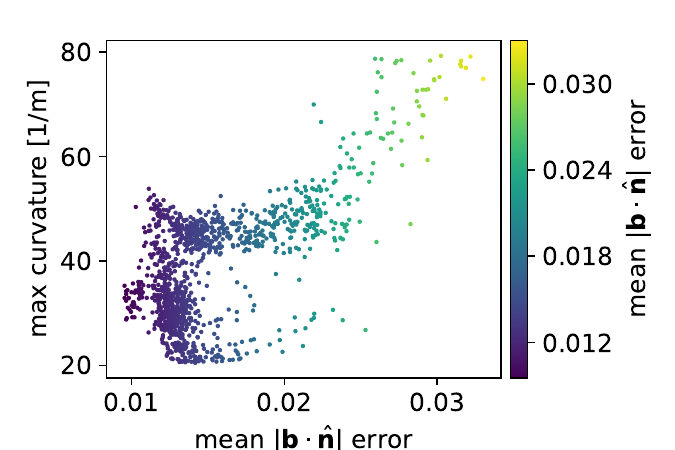}
    \end{subfigure}
    \begin{subfigure}[b]{0.32\linewidth}
        \centering
        \includegraphics[height=3cm]{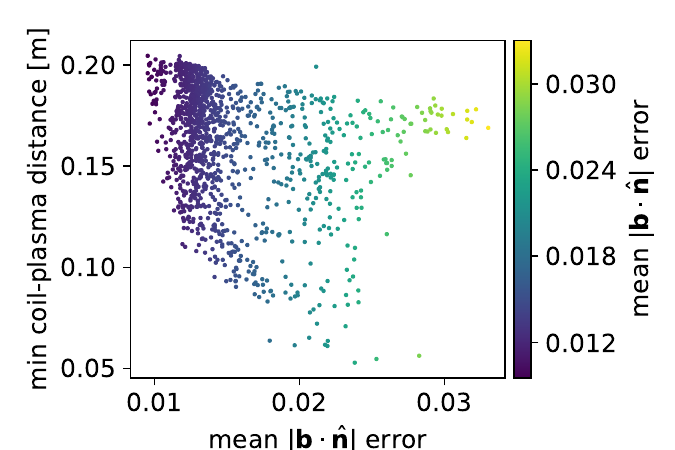}
    \end{subfigure}
    \begin{subfigure}[b]{0.32\linewidth}
        \centering
        \includegraphics[height=3cm]{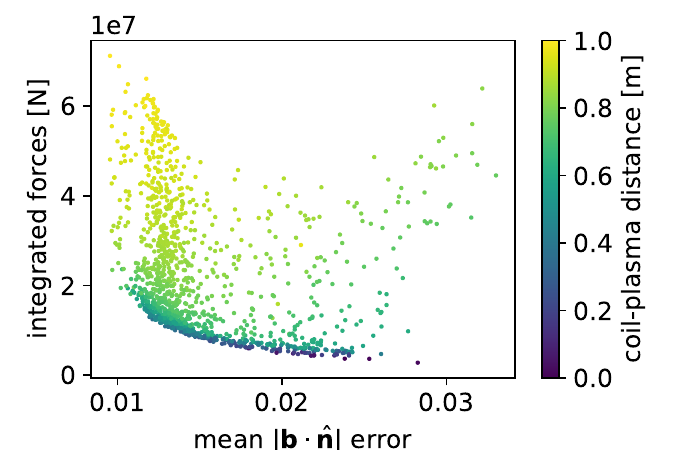}
    \end{subfigure}
    \\
    \begin{subfigure}[b]{0.32\linewidth}
        \centering
        \includegraphics[height=3cm]{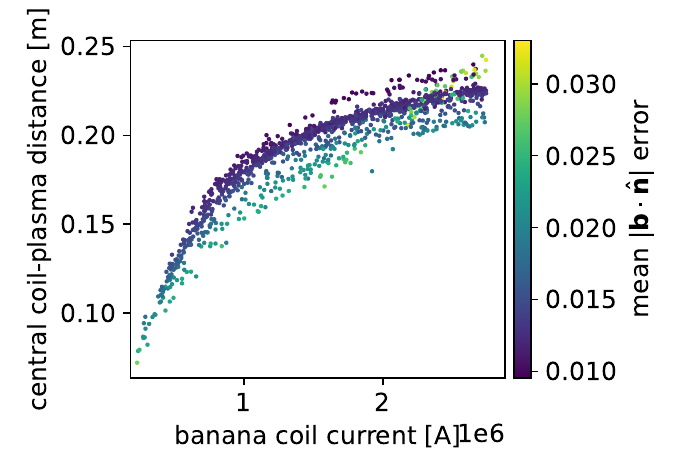}
    \end{subfigure}
    \begin{subfigure}[b]{0.32\linewidth}
        \centering
        \includegraphics[height=3cm]{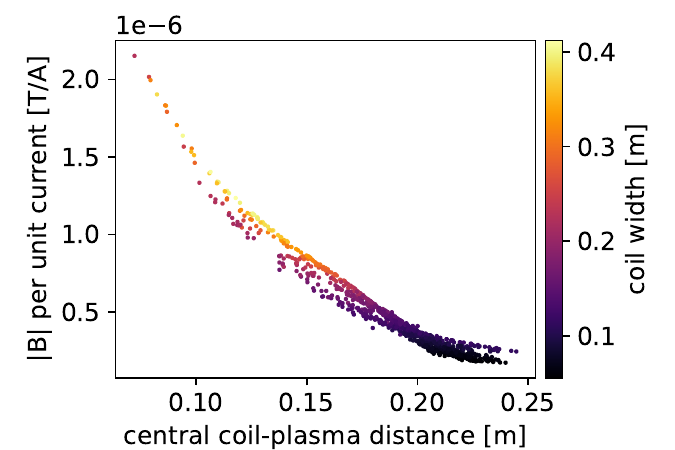}
    \end{subfigure}
    \begin{subfigure}[b]{0.32\linewidth}
        \centering
        \includegraphics[height=3cm]{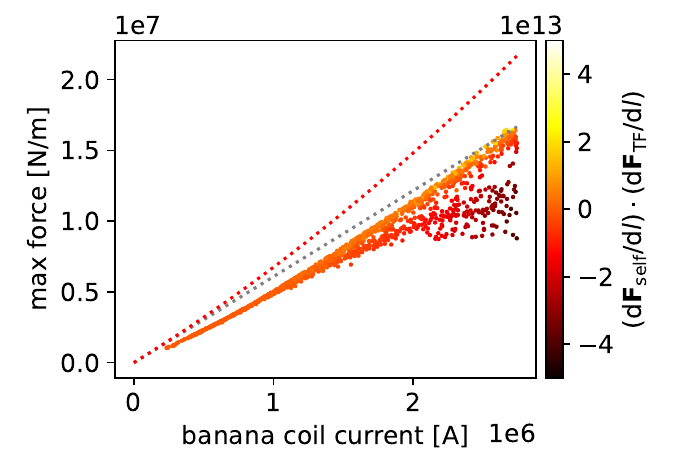}
    \end{subfigure}
    \\
    \begin{subfigure}[b]{0.32\linewidth}
        \centering
        \includegraphics[height=3cm]{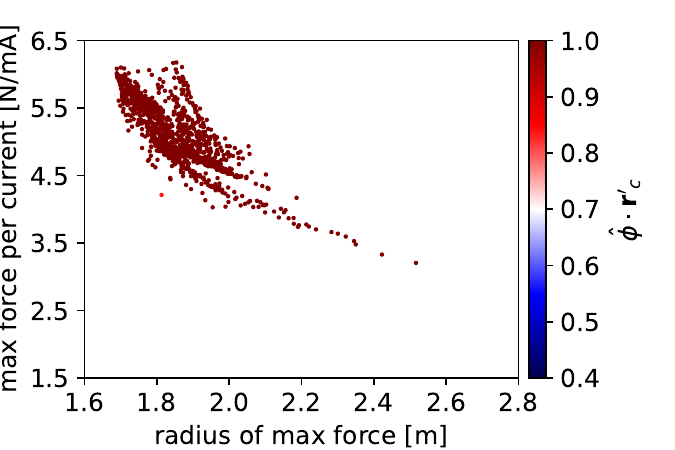}
    \end{subfigure}
    \begin{subfigure}[b]{0.32\linewidth}
        \centering
        \includegraphics[height=3cm]{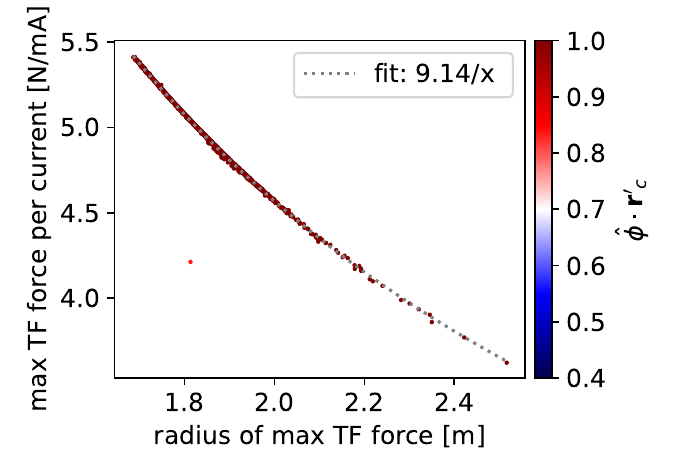}
    \end{subfigure}
    \caption{Various plots for optimised coils with zero force threshold for the H3 boundary (with $n_\text{fp}=6$, and $\iota_\text{ext} =0.05$).}
    \label{fig:data3}
\end{figure}

\suppdata{
\label{app:extradata}
In figures \ref{fig:data1}, \ref{fig:data2} and \ref{fig:data3} we present the various plots of the results from Chapter~\ref{sec:results} for the optimised coils for the H1 equilibrium with a finite force threshold of $1.75T\times I_\text{banana}$, and for the H2 and H3 equilibria, both with a zero force threshold. The plots correspond directly to those shown in Chapter~\ref{sec:results} and follow the same colour schemes and plot structure described there.}

\bibliography{banana}

\end{document}